\documentclass[a4paper,11pt]{article}
\pdfoutput=1 
             
\usepackage{defs}

\usepackage{jinstpub} 

\usepackage{comment}
\usepackage{url}
\usepackage{multirow}
\usepackage{float}
\usepackage{rotating}
\usepackage{booktabs}
\usepackage{placeins}

\usepackage{mathrsfs}
\usepackage{siunitx}
\usepackage{subfig}
\usepackage{graphicx}
\usepackage{defs}
\newbox{\bigpicturebox}
\usepackage{lineno}
\DeclareMathOperator\erf{erf}

\title{\boldmath Performance of a front-end prototype ASIC for the ATLAS High Granularity Timing Detector}

\author[a]{C. Agapopoulou,}
\author[b]{L. A.~Beresford,}
\author[c]{D. E.~Boumediene,}
\author[d]{L.~Castillo Garc\'{i}a,}
\author[e]{S. Conforti,}
\author[e]{C. de la Taille,}
\author[b]{L. D.~Corpe,}
\author[f]{M. J.~Da Cunha Sargedas de Sousa,}
\author[e]{P. Dinaucourt,}
\author[a]{A. Falou,}

\author[e]{V.~Gautam,}
\author[g]{D. Gong,}
\author[d]{C.~Grieco,}
\author[d,h]{S.~Grinstein,}
\author[b]{S.~Guindon,}
\author[i]{A.~Howard,}
\author[a]{O.~Kurdysh,}
\author[b]{E. Kuwertz,}
\author[f]{C. Li,}
\author[a]{N.~Makovec,\footnote{\label{note1}Corresponding author.}}

\author[j]{B. Markovic,}
\author[e]{G. Martin-Chassal,}
\author[k]{R.~Mazzini,}
\author[j]{C. Milke,}
\author[e]{M.~Morenas,}
\author[c]{O.~Perrin,}
\author[l]{V.~Raskina,}
\author[b]{C. Rizzi,}
\author[j]{L. Ruckman,}
\author[b]{A.~Rummler,}

\author[a]{S. Sacerdoti,}
\author[m]{G.~Saito,}
\author[e]{N. Seguin-Moreau,}
\author[a]{L. Serin,}
\author[f]{X.~Yang,}
\author[g]{J. Ye,}
\author[g]{and W. Zhou}

\affiliation[a]{Laboratoire de Physique des 2 Infinis Irène Joliot Curie (IJCLab), 15 Rue Georges Clemenceau, 91400 Orsay, France}
\affiliation[b]{Conseil Europ\'{e}en pour la Recherche Nucl\'{e}aire (CERN), Esplanade des Particules 1, CH-1211 Meyrin, Switzerland}
\affiliation[c]{Laboratoire de Physique de Clermont-Ferrand (LPC), Universite Clermont Auvergne, Campus Universitaire des Cézeaux, 4 Avenue Blaise Pascal, 63178 Aubière Cedex, France}
\affiliation[d]{Institut de F\'{i}sica d'Altes Energies (IFAE), The Barcelona Institute of Science and Technology (BIST), \\Carrer Can Magrans s/n, Edifici Cn, Campus UAB, E-08193 Bellaterra (Barcelona), Spain}
\affiliation[e]{Omega, Ecole Polytechnique, Palaiseau, France}
\affiliation[f]{Department of Modern Physics and State Key Laboratory of Particle Detection and Electronics, University of Science and Technology of China (USTC), 96  JinZhai Road Baohe District, Hefei, Anhui, 230026, China}
\affiliation[g]{Southern Methodist University, Department of Physics, Dallas, Texas}
\affiliation[h]{Instituci\'{o} Catalana de Recerca i Estudis Avan{\c c}ats (ICREA), Passeig de Llu\'{i}s Companys, 23, 08010 Barcelona, Spain}

\affiliation[i]{Jo\v{z}ef Stefan Institute (JSI), Jamova cesta 39, 1000 Ljubljana, Slovenia}
\affiliation[j]{SLAC National Accelerator Laboratory, Stanford CA; United States of America}
\affiliation[k]{Academia Sinica, 128, Section 2, Academia Road, Nangang District, Taipei City, Taiwan 115}
\affiliation[l]{Laboratoire de Physique Nucl\'{e}aire et de Hautes Energies (LPNHE), Sorbonne Universit\'{e}, Universit\'{e} de Paris, CNRS/IN2P3, Paris, France}
\affiliation[m]{Instituto de Fisica, Universidade de Sao Paulo, Sao Paulo, Brazil}

\emailAdd{makovec@cern.ch}

\abstract{
This paper presents the design and characterisation of a front-end prototype ASIC for the ATLAS High Granularity Timing Detector, which is planned for the High-Luminosity phase of the LHC.
This prototype, called ALTIROC1, consists of a 5$\times$5-pad matrix and contains the analog
part of the single-channel readout (preamplifier, discriminator, two TDCs and SRAM).
Two preamplifier architectures (transimpedance and voltage) were implemented and tested. 
The ASIC was characterised both alone and as a module when connected to a 5$\times$5-pad array of LGAD sensors.
In calibration measurements, the ASIC operating alone was found to satisfy the technical requirements for the project, with similar performances for both preamplifier types. In particular, the jitter was found to be 15$\pm$1~ps (35$\pm$1~ps) for an injected charge of 10~fC (4~fC).
A degradation in performance was observed when the ASIC was connected to the LGAD array. This is attributed to digital couplings at the entrance of the preamplifiers. 
When the ASIC is connected to the LGAD array, the lowest detectable charge increased from 1.5~fC to 3.4~fC. As a consequence, the jitter increased for an injected charge of 4~fC. Despite this increase, ALTIROC1 still satisfies the maximum jitter specification (below 65~ps) for the HGTD project.
This coupling issue also affects the time over threshold measurements and the time-walk correction can only be performed with transimpedance preamplifiers.
Beam test measurements with a pion beam at CERN were also undertaken to evaluate the performance of the module. The best time resolution obtained using only ALTIROC TDC data was 46.3$\pm$0.7~ps for a restricted time of arrival range where the coupling issue is minimized.
The residual time-walk contribution is equal to 23~ps and is the dominant electronic noise contribution to the time resolution at 15~fC.

}

\begin{document}
\maketitle
\flushbottom
\tableofcontents
\newpage

\section{Introduction}
\label{sec:intro}
The large increase in the number of interactions per bunch crossing (known as pileup) is one of the main experimental challenges for the High-Luminosity LHC (HL-LHC~\cite{hllhc}) physics program.
The High Granularity Timing Detector (HGTD)~\cite{HGTD} will contribute to pileup mitigation in the forward region of the ATLAS detector~\cite{ATLAS} by exploiting high-precision timing information to
distinguish between collisions occurring close in space but well-separated in time.
In addition, the number of collected hits being proportional to the luminosity, it will provide an instantaneous measurement of the luminosity, reading this information at 40 MHz. 
The HGTD will use Low Gain Avalanche Diode (LGADs)~\cite{lgad,4Dtracking} sensors, which are designed to provide a most probable charge value of 4fC for a minimum ionizing particle at the largest radiation level seen by the detector (which is 2.5$\times$10$^{15}$ n$_{\mathrm{eq}}$/cm$^2$ and 2.5 MGy). 
The typical number of hits per track in the detector was optimized so that the target average
time resolution per track for a minimum ionising particle is 30~ps at the start of
the HGTD lifetime, increasing to 50~ps at the end of HL-LHC operation. 

Each LGAD sensor pixel is read out by a front-end ASIC channel. This ASIC, called ALTIROC (ATLAS LGAD Time Read Out Chip), is one of the key elements in the design of this new detector concept. This ASIC must match the excellent timing performance of the LGAD.
 For unirradiated LGAD sensors, the electronic jitter for an input charge of 10~fC is required to be smaller than 25~ps. At the maximum fluence, the requirement is relaxed due to the reduced sensor gain: the electronic jitter for an input charge of 4~fC is required to be smaller than 65~ps. The time-walk effect (after correction) and time-to-digital converter contributions should be smaller than 10~ps. Since the most probable charge value is 4~fC at the maximum fluence and considering the width of Landau distribution, the lowest detectable charge should be at most 2~fC to have an efficiency to detect a hit above 95\%.

 The ALTIROC design is such that the signal is amplified before entering a fast discriminator. A time-to-digital converter (TDC) system provides the time of arrival (TOA) with respect to the 40 MHz clock if the rising edge of the discriminator arrives within a 2.5~ns window centred on the LHC bunch crossing. A second TDC, using both the rising and falling edges of the discriminator, provides a time over threshold (TOT) used as an estimate of the signal amplitude to correct for the time-walk effect offline. 
 The performance of the first prototype version (ALTIROC0), which contained only the preamplifier and the discriminator, was described in Ref.~\cite{altiroc0}. In this paper, the design and performance of a second prototype version (ALTIROC1) are presented.
This second version consists of a $5\times 5$-pad matrix instead of $2\times 2$-pad matrix, in which two TDCs and an SRAM were added to the single-channel readout.
Two preamplifier architectures were implemented: a transimpedance preamplifier (TZ) for the 15 first channels and a voltage preamplifier (VPA) for the remaining 10 channels.

The ALTIROC1 design is described in section~\ref{sec:design}, followed by a section on the readout system used to characterise the devices (section~\ref{sec:devices}). The results are presented in two different sections, the first contains the results obtained on a test bench with a calibration signal (section~\ref{sec:testbench}) and the second contains results obtained in beam tests (section~\ref{sec:testbeam}).

\section{Design}
\label{sec:design}
\subsection{Preamplifier} 

High-frequency (GHz) preamplifiers are essential for picosecond time measurements as the jitter scales with the noise divided by the signal slope (or bandwidth). As shown in Ref.~\cite{altiroc0}, the rise time also depends on the sensor current duration (typically 500 ps) as this current initially gets integrated with the sensor capacitance. The optimum preamplifier speed is obtained by balancing risetime against sensor current duration (a faster preamplifier just adds electronic noise whereas a slower one degrades the jitter).

\begin{figure}[hptb]
\centering
\includegraphics[width=0.6\textwidth,angle=0]{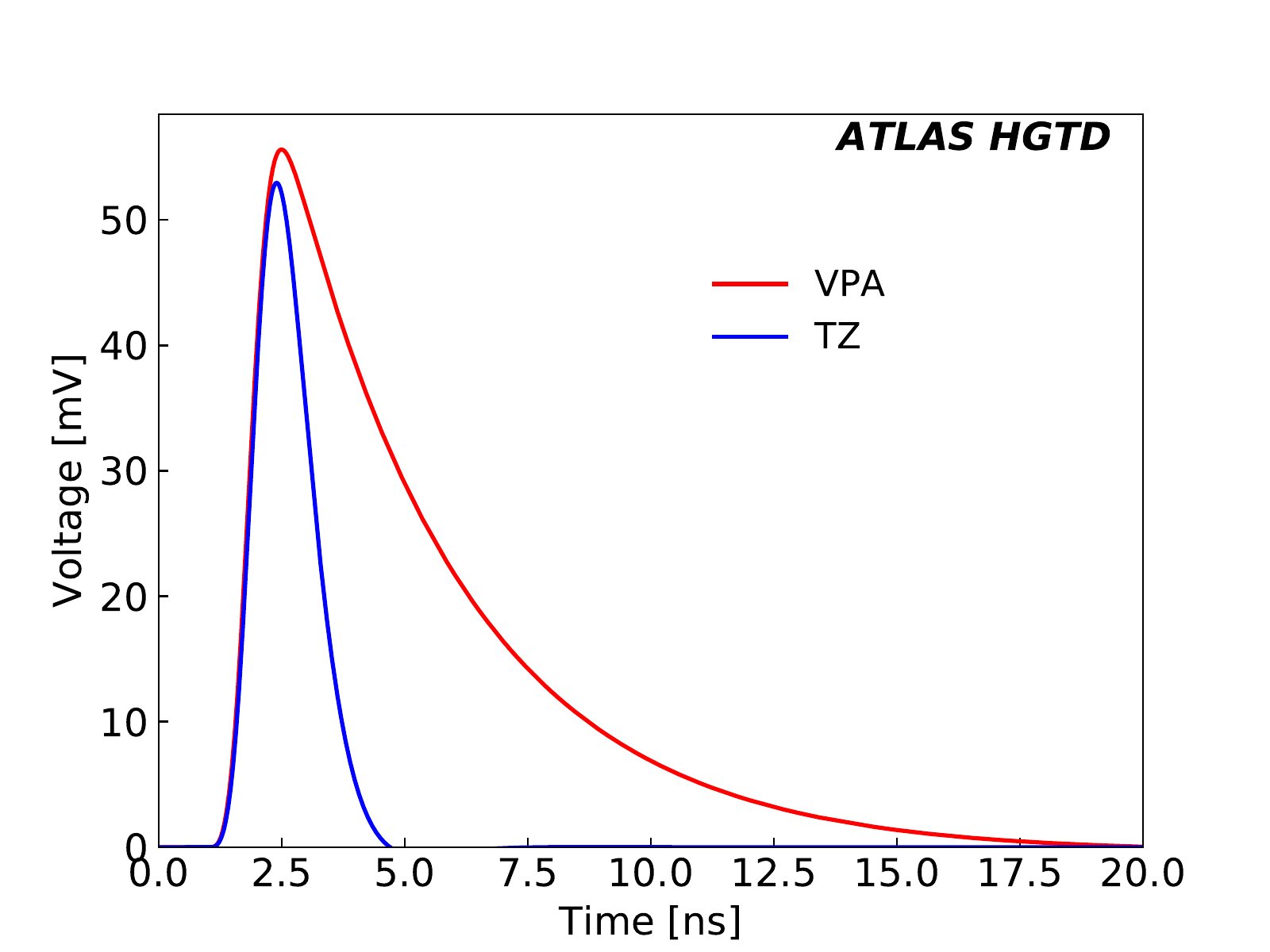}
\caption{Output amplitude as a function of time for voltage and transimpedance preamplifiers for an LGAD signal. The voltage preamplifier is labelled VPA while the transimpendence preamplifier is labelled TZ.
\label{fig:simu1}}
\end{figure}

In early studies, commercial or custom-made discrete radio frequency (RF) voltage amplifiers were used to read out the LGAD sensors. In contrast, for the ASIC, two architectures were tested: voltage-sensitive and current- or transimpedance-sensitive preamplifier configurations were made available, differing in resistor feedback. Indeed, low-feedback resistors yield small input impedance and current-sensitive preamplifiers, whereas high-feedback resistors lead to voltage-sensitive preamplifiers. The input impedance has no impact on the output risetime but changes the decay time, potentially affecting the occupation time. In principle, the voltage-sensitive configuration exhibits lower parallel noise (larger RF) but the longer signal duration makes it more sensitive to low-frequency noise, such as the digital noise discussed below. Moreover, the 40 MHz repetition rate at the LHC precludes a decay time longer than a few tens of nanoseconds and therefore the resulting ``voltage preamplifier'' more closely resembles another transimpedance preamplifier, but the naming scheme is maintained to distinguish the two configurations. The ``voltage preamplifier'' uses a 25 k$\Omega$ feedback resistor (RF) giving an input impedance of $\sim$ 1 k$\Omega$, whereas the ``transimpedance preamplifer'' uses 4k$\Omega$ giving $\sim$300 $\Omega$. The corresponding output waveforms can be seen in Figure~\ref{fig:simu1}.

\begin{figure}[htbp]
\centering
\includegraphics[width=0.8\textwidth]{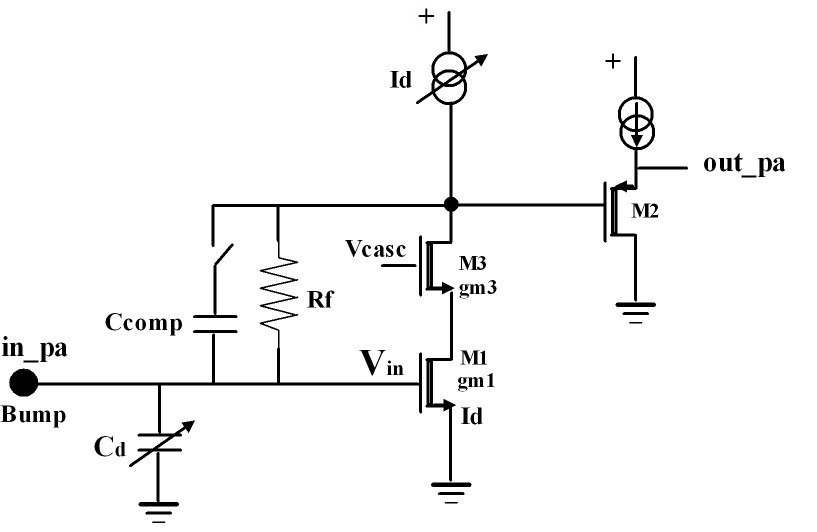}\label{fig:TZ}
\caption{Preamplifiers schematics. \label{fig:PAschematics}}
\end{figure}

Both preamplifiers are built around a cascoded common source NMOS amplifier to ensure high bandwidth (see Figure \ref{fig:PAschematics}). The drain current ($I_d$) of the input transistor is adjustable between 200 $\mu$A and 1 mA. The transistor size is optimized to operate close to weak inversion while keeping its capacitance small compared to that of the sensor. The operating current is chosen to minimize the series noise while not dissipating too much power ($< 2.25$ mW/ch for the analog part). A PMOS follower is added to isolate the load from the discriminator. The total preamplifier power consumption is 0.85 mW using a nominal current $I_d = 600 \mu$A in the input transistor. A bank of seven capacitors (from 0 to 3.5 pF) can be connected by slow control to the preamplifier input to emulate the sensor capacitance when measuring the ASIC alone. They are not used when the ASIC is connected to the LGAD sensor array.

\subsection{Discriminator}

The measurement of the TOA of the particles is performed by a discriminator that follows the preamplifier.
The discriminator rising edge position provides the TOA. Combining this information with the falling edge position,  provides the TOT.
To ensure a jitter smaller than 10~ps, the discriminator is built around a high-speed leading edge architecture with hysteresis to avoid re-triggering effects.
Two differential stages with small input transistors are used to ensure a large gain and a large bandwidth ($\sim$\SI{0.7}{\giga\hertz}).
A 10-bit DAC (with a step size of 0.4~mV) is used to set a common discriminator threshold ($V_{\mathrm{th}}$) to all channels although 9 bits would have been sufficient.
An additional 7-bit DAC (with a step size of 0.8~mV) allows threshold corrections to be made individually for each channel to compensate for differences amongst them or for different values of sensor leakage current.
The TOT obtained from the width of the discriminator is used to perform time-walk correction of the TOA measurement.
Each discriminator output is sent to a sampling cell to generate a ``Hit Flag'' bit, that is equal to 1 in case of a hit or to 0 in case of no hit.
The discriminator's power consumption is slightly less than 0.4 mW.

\subsection{Time-to-digital converter}

The target quantisation step of the TDC for the TOA is \SI{20}{\pico\second}, and is below the gate-propagation delay in \SI{130}{\nano\metre} technology, thus the Vernier delay line configuration is employed.
This configuration consists of two lines, each composed of a series of delay cells implemented as differential shunt-capacitors, controlled by a voltage signal that determines their delay.
The timing resolution is determined by the difference in the delays of the cells in each line.
The TOA will be measured within a \SI{2.5}{\nano\second} window centred at the bunch-crossing.
An internal phase shifter is used to align events within the 2.5~ns acceptance window.
The hits have a time dispersion with an RMS of about \SI{300}{\pico\second}, so that such a window aligned with a precision of \SI{100}{\pico\second} contains all the hits.
The maximum conversion time for a \SI{2.5}{\nano\second} range must be below \SI{25}{\nano\second} so that hits happening in the following bunch crossing can be converted.

\begin{figure}[htbp]
\centering
\includegraphics[width=1.0\textwidth]{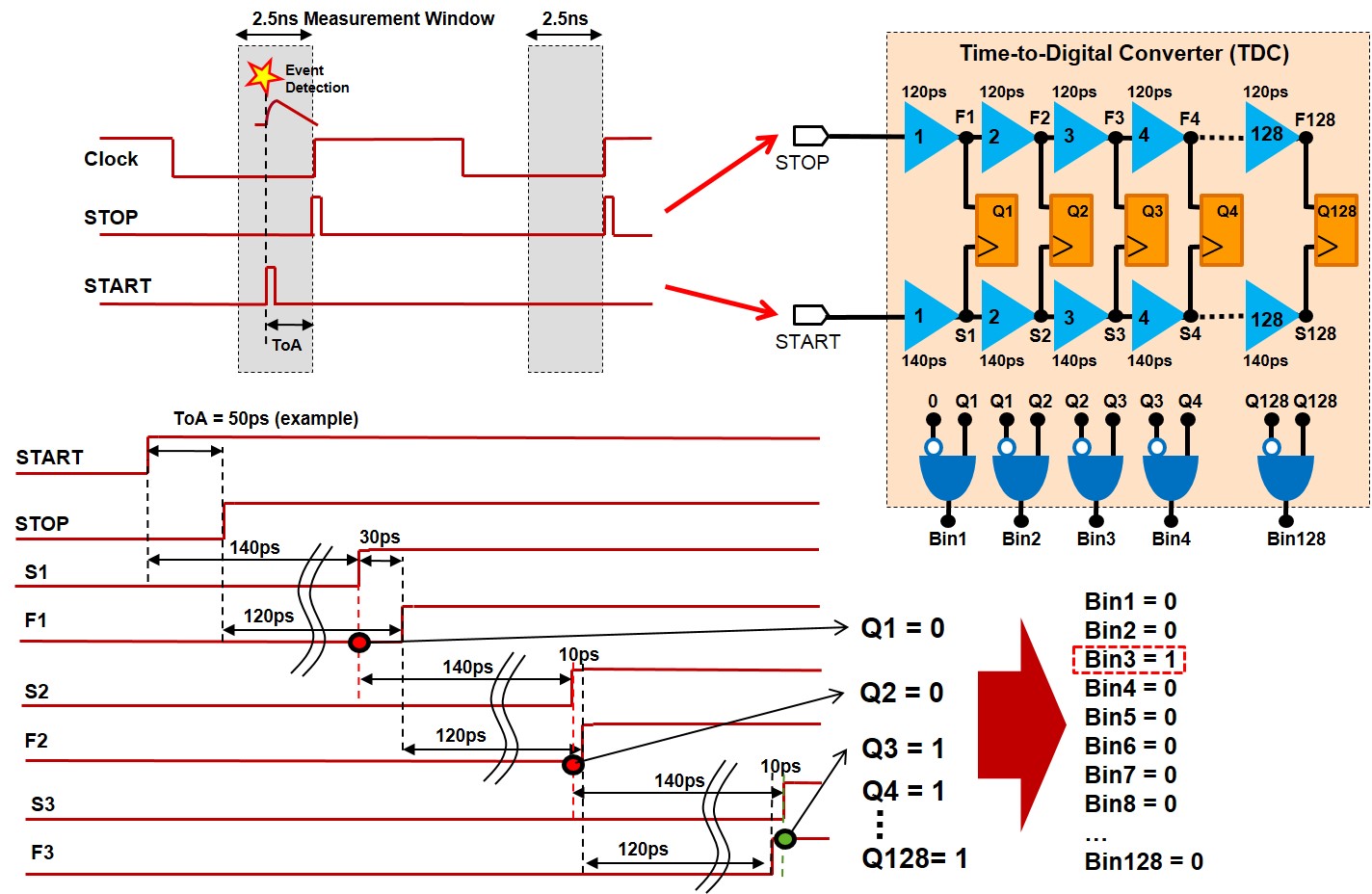}
\caption{
  Graphical representation of the working principle of the TDC.
  The drawing on the top left shows how the START and STOP signals are generated, the first with the discriminator output upon event
  detection, the second corresponding to the next clock edge. The gray area indicates the \SI{2.5}{\nano\second} detection window.
  On the top right, the schema represents the TDC, with the ``slow'' delay line (\SI{140}{\pico\second} cells) that propagates the START signal, and the
  fast delay line (\SI{120}{\pico\second} cells) in which the STOP signal is propagated. The difference between delays defines the bin.
  After each cell the signals are compared (Q1, Q2, ...), and the bin number provides the converted measurement.
}
\label{fig:TDCexplained}
\end{figure}

A graphical representation of the working principle of the TDC can be found in Figure~\ref{fig:TDCexplained}.
In the slow line, the control voltage fixes the delay of each cell to \SI{140}{\pico\second}, while on the fast line it fixes it to \SI{120}{\pico\second}.
The START signal (rising edge of the discriminator) enters the slow delay line while the STOP signal (next rising edge of the 40MHz clock) enters the fast delay line.
Although initially the START signal is ahead of the STOP one, each delay-cell stage brings them closer by an amount equal to the difference between the slow and fast cell delays, i.e.~\SI{20}{\pico\second}.
The number of cell stages necessary for the STOP signal to surpass the START signal represents the result of the time measurement with a quantisation step of \SI{20}{\pico\second}.
A cyclical structure is employed to reduce the number of cells per line and results in a smaller occupied area.
Since the time measurement is initiated only upon signal detection (instead of at each time-measurement window), the reverse START-STOP scheme is used as a power-saving strategy.
The conversion time of a \SI{2.5}{\nano\second} input time interval is \SI{21}{\nano\second}, finishing before the next bunch crossing.

The TOT TDC provides a 9-bit digitization of the discriminator width, on a \SI{20}{\nano\second} range. It uses an additional coarse delay line made of \SI{160}{\pico\second} delay cells to extend the measurement range to \SI{20}{\nano\second}, while a Vernier delay line provides the requested fine resolution of \SI{40}{\nano\second}. The START and STOP signals are given by the rising edge and by the falling edge of the discriminator respectively.

As mentioned before, the delay cells of both TDCs are implemented as differential shunt-capacitor voltage-controlled delay cells. Their delay is set by a control voltage ($V_{\mathrm{ctrl}}$) that controls the load of the cell.
Three control voltages are necessary to control the three delay lines used in the TDCs: $V_{\mathrm{ctrl\_fast}}$ to set the cell delay of fast cells to the desired value of 120 ps, $V_{\mathrm{ctrl\_slow}}$ to set the cell delay of slow cells to 140 ps for slow cells and $V_{\mathrm{ctrl\_coarse}}$ to set the cell delay of slow cells to 160 ps. These control voltages are generated by three delay-locked loops (DLLs) located in the periphery of the ASIC and built around a classical architecture (phase comparator and a charge pump current). Additional open-loop per-channel trimming is present to minimize dispersion between channels due to cells mismatches.
Each DLL uses the very same delay cells as those used in the corresponding controlled delay lines: this ensures that the TDC steps don’t vary either with PVT (Process Voltage Temperature) parameters or under irradiation. 
A compensation system is integrated for each DLL in case if improper locking. This system consists of adding or removing a tuneable current into the charge pump capacitor. 
On the test bench some of the DLLs exhibited locking issues, so this compensation system was used for the measurements described in this paper.


The TDC power consumption is dependent on the time-interval being measured.
For the TOA TDC using \SI{2.5}{\nano\second} (full dynamic range), the average power consumption over the \SI{25}{\nano\second} measurement period is about \SI{5.2}{\milli\watt}.
It is only \SI{3.5}{\milli\watt} for a time-interval equal to half the dynamic range.
Thanks to the reverse START-STOP operation, the power consumption of the TDC is much lower in the absence of a hit over threshold.
This results in an average power consumption per channel of \SI{1.1}{\milli\watt} for both TDCs, assuming a time interval uniformly distributed (\SI{1.25}{\nano\second} average) and a maximal channel occupancy of 10\%.

\subsection{SRAM}
A memory is integrated for each channel. It is implemented as a circular memory (SRAM) to continuously store data, with a depth of 10 $\mu$s. It is built around a standard 6T cell configuration and it is written at a rate of 40 MHz. However, for power saving purposes, only the Hit Flag bit generated by the discriminator sampling cell is written if no hit occurred. Conversely, if there is a hit, the 16 bits of time measurement information (7 bits for the TOA and 9 bits for the TOT) as well as the Hit Flag bit are written. The FIFO is read for each L1 signal arrival and the readout data are then serialized before being sent to the acquisition system.

\subsection{Calibration: charge and digital injection}
\label{sec:pulser}

\begin{figure}[hptb]
\centering
\includegraphics[width=1.0\textwidth,angle=0]{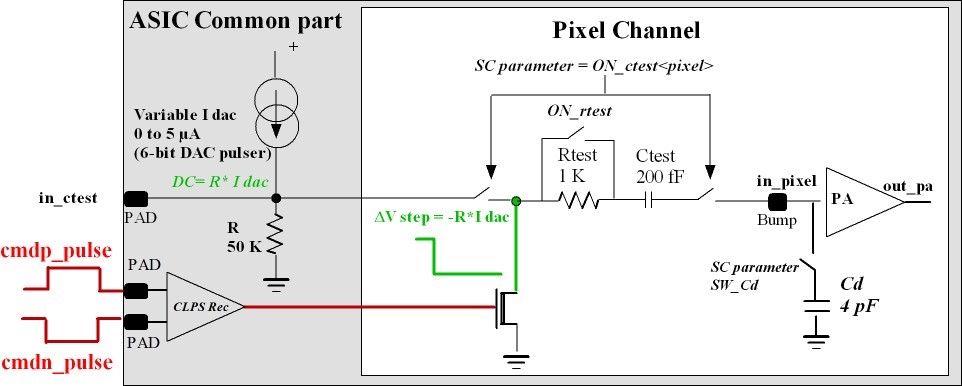}
\caption{  
Pulser schematics.
}
\label{fig:pulser}
\end{figure}

An internal pulser, common to all channels, is integrated to inject charge, producing a signal mimicking one from an LGAD sensor.
The pulser consists of a programmable DC current, which is tuneable with an internal 6-bit DAC,
that flows continuously through a 50 kΩ resistor ($R$) until it is interrupted by a command pulse, cmd$_{\textrm{pulse}}$, that shorts
the resistor to ground (see Figure~\ref{fig:pulser}).
A voltage step (V$_{step}$) equal to $-R\times I_{DAC}$,
is then generated and sent through the internal 200~fF test capacitors (C$_{test}$) of the selected channel.
The input charge ($Q_{inj}$) is equal to $C_{test}\times V_{step}$ and the dynamic range goes from 0 to ~250 mV or 0~fC up to $\sim$40~fC (step $\sim$ 0.6 fC).
A resistor R$_{test}$ can be added in series with the C$_{test}$ capacitor in order to slow down the injected calibration signal and mimic the rising edge (around 600~ps).
The absolute value of C$_{test}$ and R$_{test}$ are known within 10\% and the relative value between channels is within 1\%.
The command pulse is synchronous with the 40~MHz clock and its relative phase can be moved through slow control parameters by steps of 10~ps to perform the calibration of the TOA bin.

To study the TDC performance specifically, but also to perform the calibration of the TOT bins, a digital injection (called external discriminator) is used. In this mode the discriminator output is bypassed, and a discriminator-like signal is injected in the TDC. Both the start time and the width of this signal can be tuned with a 10~ps step by slow control parameters.

\subsection{Layout} 

\begin{figure}[htbp]
\centering
\subfloat[]{\includegraphics[width=0.4\textwidth,angle=0]{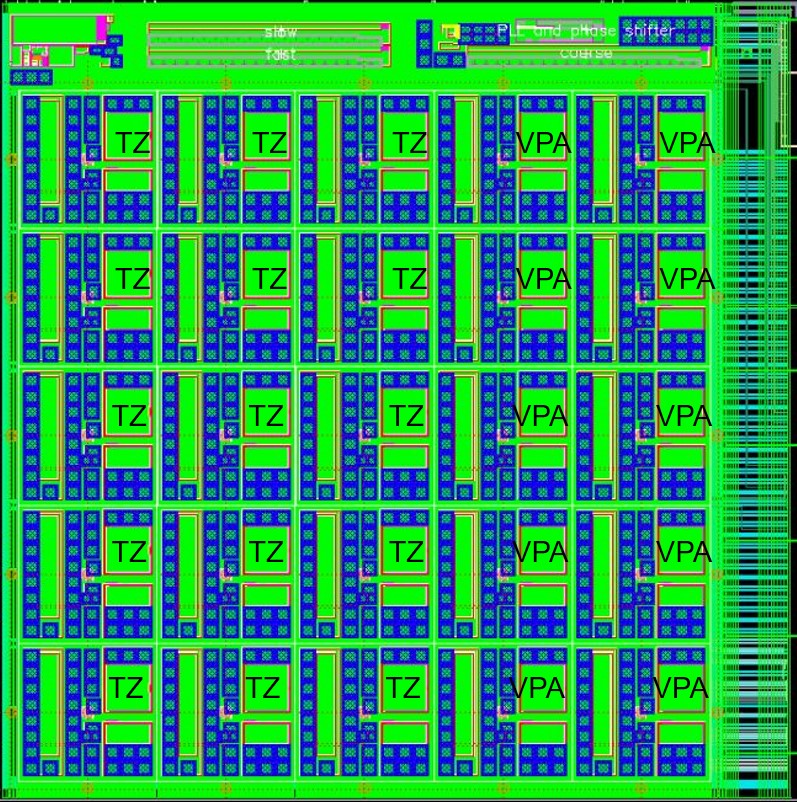}\label{fig:layout}}
\hspace*{0.2cm}
\subfloat[]{\includegraphics[width=0.4\textwidth,angle=0]{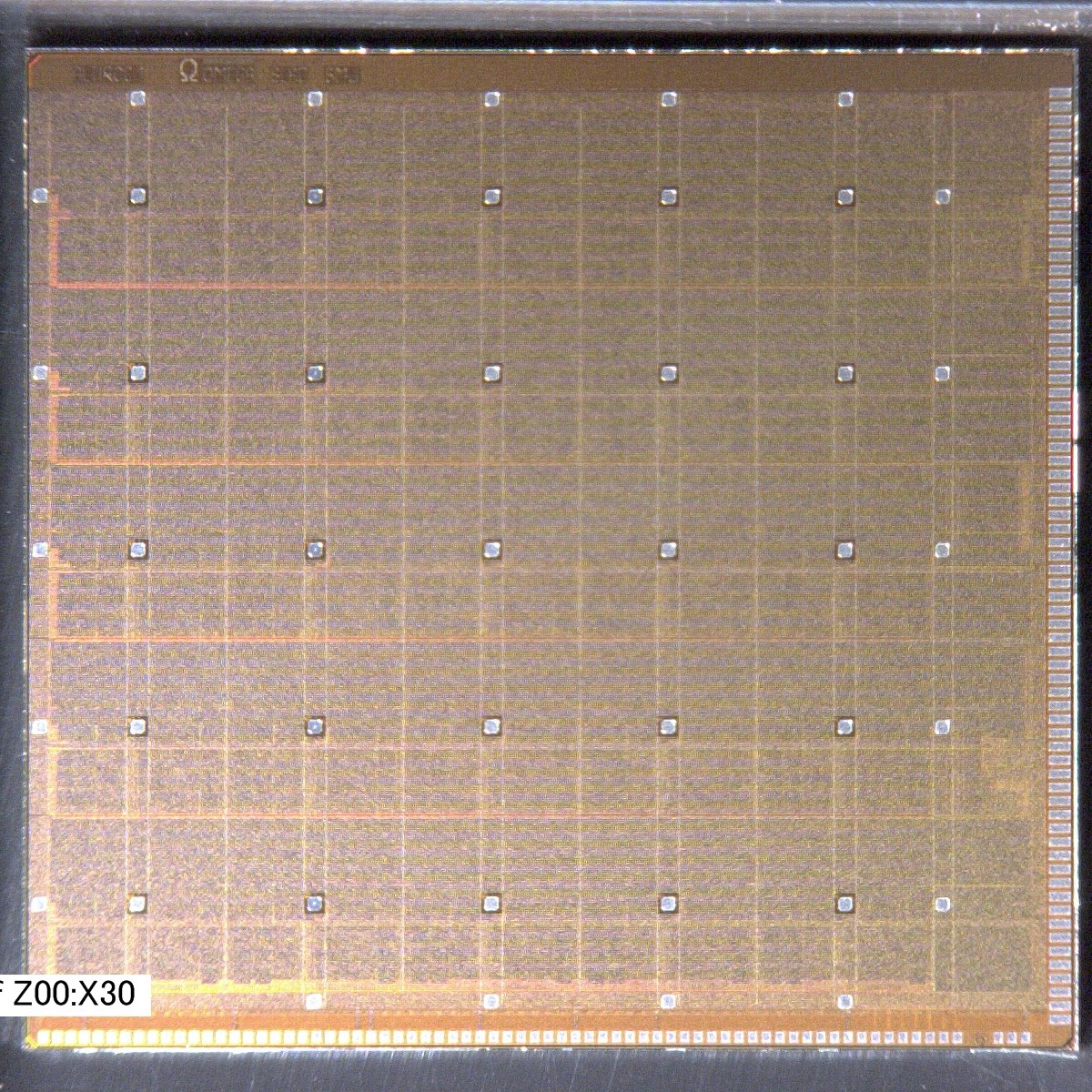}\label{fig:ALtiroc1_die}}
\caption{ALTIROC1 layout (a) and dice (b).}
\end{figure}

A prototype ASIC with 25 channels has been designed using CMOS 130~nm technology. It consists of 15 transimpedance preamplifier channels that equip the first three columns, and 10 voltage preamplifier channels that equip the last two columns, as can be seen in Figure~\ref{fig:layout}. 
Each channel is made of a preamplifier followed by a discriminator, 2 TDCs and an SRAM with a depth of 10 $\mu$s.
The size of the chip is 7.6 mm $\times$  7.7 mm to accommodate the bump bonding to a sensor array with a 1.3 mm  $\times$ 1.3 mm pad size.


To understand the performance of the ASIC, an analog probe was integrated inside the prototype ASIC.
This probe allows the preamplifier signal to be read out by an oscilloscope.
When this probe is enabled, the preamplifier output is not only sent to the discriminator
but also to an amplifier allowing signal to be monitored with an oscilloscope.
In a similar way, a digital probe allows the output of the discriminator,
that is sent to the TDC, to be monitored with an oscilloscope.
A summary of the main characteristics of ALTIROC1 can be found in table~\ref{tab:design}.

\begin{table}[ht]
    \centering
    \begin{tabular}{|c|c|}
        \hline
        ALTIROC1 size & 7.6 mm $\times$  7.7 mm  \\
        \hline
        ALTIROC1 pad size & 1.3 mm  $\times$ 1.3 mm \\
        \hline
        Preamplifier power consumption & 0.85~mW \\
        \hline
        Discriminator power consumption & 0.4~mW \\
        \hline
        TOA and TOT TDC power consumption & 1.1~mW \\
        \hline
          \end{tabular}
    \caption{Main characteristics of ALTIROC1.}
    \label{tab:design}
\end{table}

\section{Readout system and devices}
\label{sec:devices}
\begin{figure}[htbp]
\centering
\subfloat[]{\includegraphics[width=0.515\textwidth]{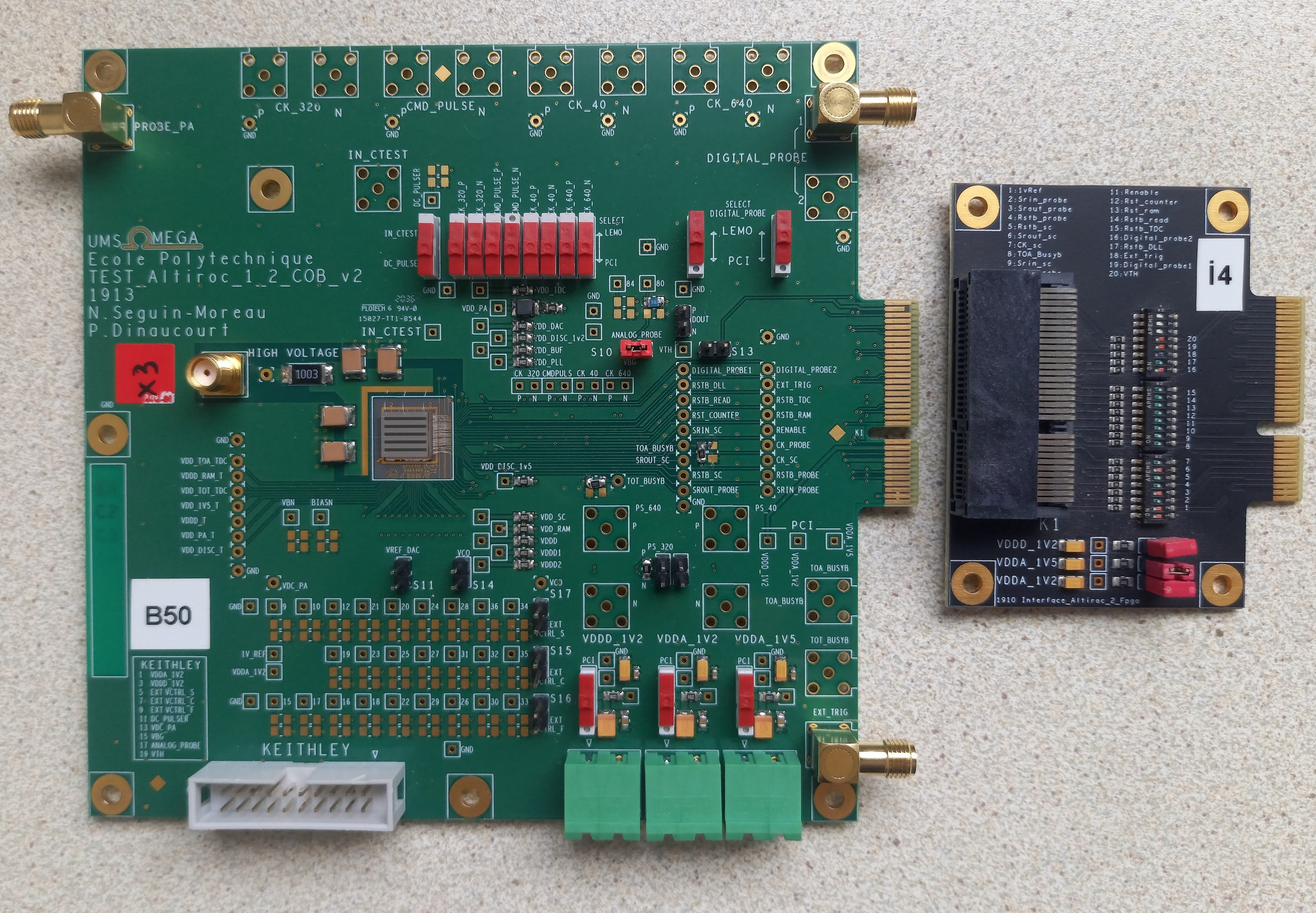}\label{fig:B50}}
\hspace*{0.2cm}
\subfloat[]{\includegraphics[width=0.485\textwidth]{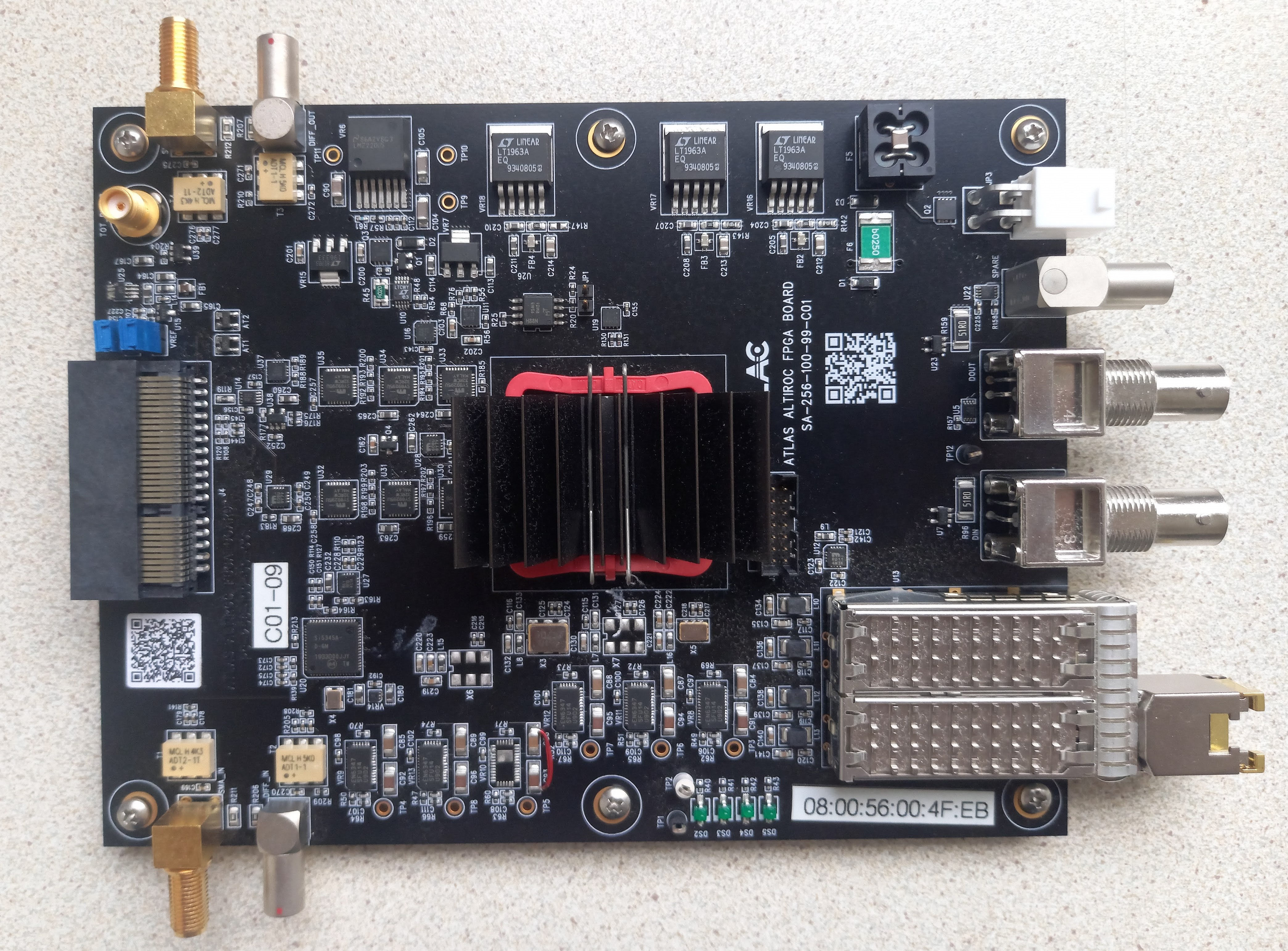}\label{fig:FPGA}}
\caption{
Photograph of an ALTIROC1 read-out board together with
the small noise-filtering interface board (a) and an FPGA board (b). On the ALTIROC1 board, the ASIC is bump-bonded to an LGAD.
}
\end{figure}

The custom ASIC read-out board of ALTIROC1 is shown in Figure~\ref{fig:B50}. 
The ASIC is mounted
in the dedicated central region of the board. 
An L-shaped high voltage (HV) pad is positioned next to it, providing space
for the HV wire bonds needed for the sensors.
To understand the performance of the ASIC prior to the digitization step, two
analogue probes, one for the preamplifier and one for the discriminator output, are integrated.
The board also contains various probes of the DC voltages that configure the delay values of
the TDC cells.

A separate board was designed to integrate the custom
FPGA, the powering units and a clock generator. 
A photograph of this board can be seen in Figure~\ref{fig:FPGA}. 
The FPGA is responsible for the transmission of clocks and slow control bits to
the ASIC as well as the digital data acquisition. 
Dedicated firmware and software were developed for the characterisation of ALTIROC1~\cite{fwsw}.
A PCIe cable ensures the interconnection of the two boards. 
The system can be
operated either in self-trigger mode when the ASIC registers a hit, or as a ``slave'' by receiving an external trigger. 
It can also provide a trigger output, allowing for synchronization with
other devices.

To reduce the noise contributions of various powering and
digital signals to the ASIC, a small interface board, shown in Figure~\ref{fig:B50}, was designed and integrated in the system. 
This board is connected between the ASIC
board and the PCIe cable end and allows to switch various signals and add RC filters through
footprints on the back side. 

For the measurements presented in this paper, two boards were used. The first had only an ASIC mounted on it and will be referred as the ``ASIC board'' hereafter.
The second one, shown in Figure~\ref{fig:B50}, had an ASIC bump-bonded to a sensor and will be referred as the ``sensor board''.
For this latter board, an HPK2 W42 5$\times$5-pads sensor produced by Hamamatsu Photonics was used.
HPK2 is the second sensor HPK production focused on the research and development for the ATLAS and
CMS timing detectors~\cite{ferrero}. This production has a deep and narrow multiplication layer
implanted in a high-resistivity bulk. Four splits of p-gain dose are implanted. HPK2 W42 belongs to the fourth split with the lowest doping level.
The interconnection between the ASIC and the sensor was performed with the flip chip method after under-bump metallization deposition using 115 $\mu$m thick SnAg bumps.



\section{Test bench results}
\label{sec:testbench}
This section presents the characterisation of ALTIROC1 with a test bench.
For the ASIC board, a capacitor was
connected through a programmable switch to the preamplifier input,
mimicking the LGAD sensor capacitance
and thus allowing the study of the device's performance as a function of the detector 
capacitance $C_d$. The capacitance was tuneable from \SIrange{0}{3.5}{\pico\farad} with a step of \SI{0.5}{\pico\farad} and the maximal value was always used to be as close as possible to the detector capacitance of an LGAD.
All the measurements were performed with only one channel activated, meaning the preamplifiers, discriminators, TDC and SRAM of all other channels were turned off.
When an ASIC was connected to a sensor, the latter was always operated at a bias voltage high enough to ensure its full depletion.

%

\begin{figure}[htbp]
\centering
\subfloat[]{\includegraphics[width=0.5\textwidth]{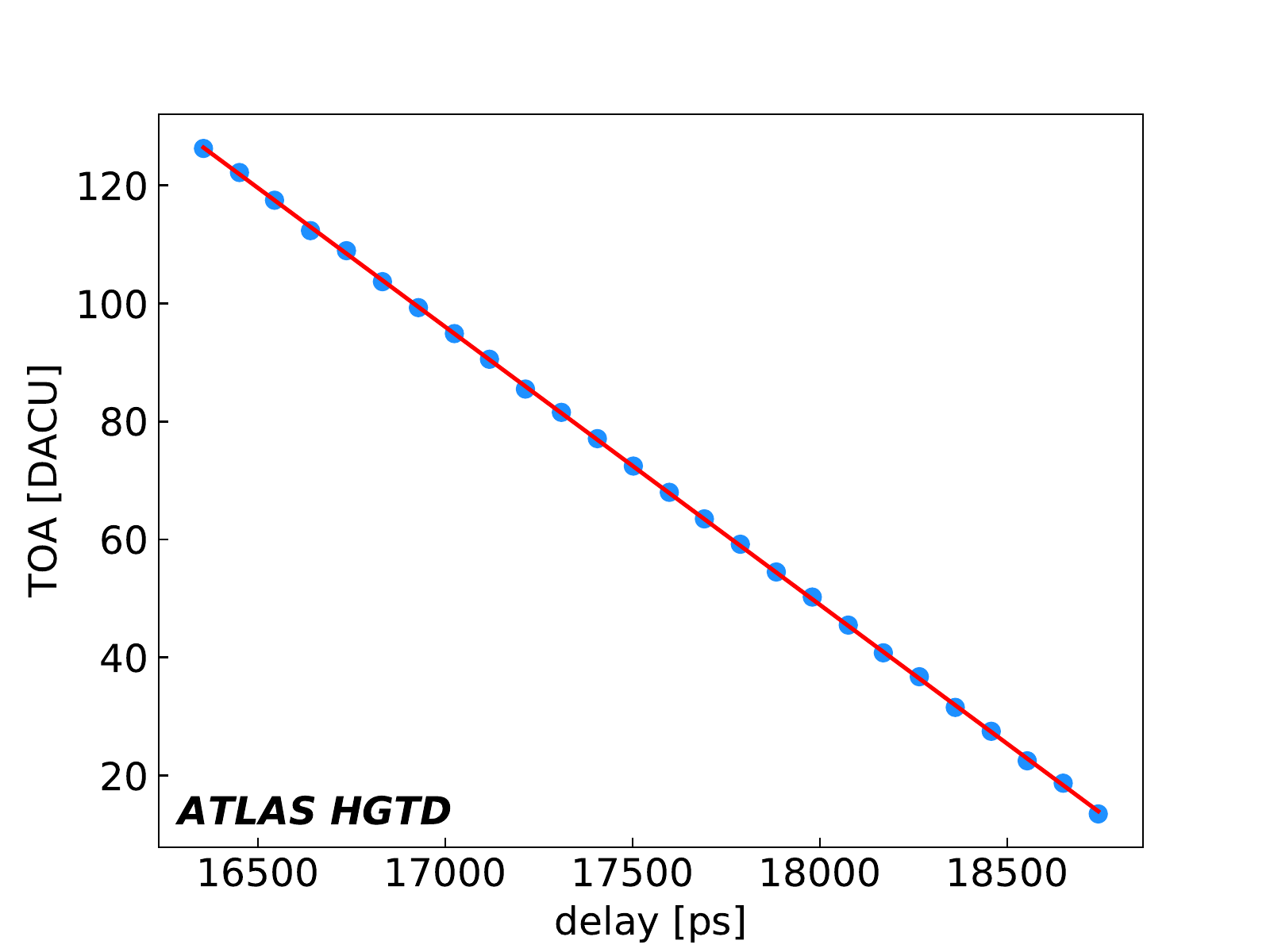}\label{fig:lsbtoa}}
\hspace*{0.2cm}
\subfloat[]{\includegraphics[width=0.5\textwidth]{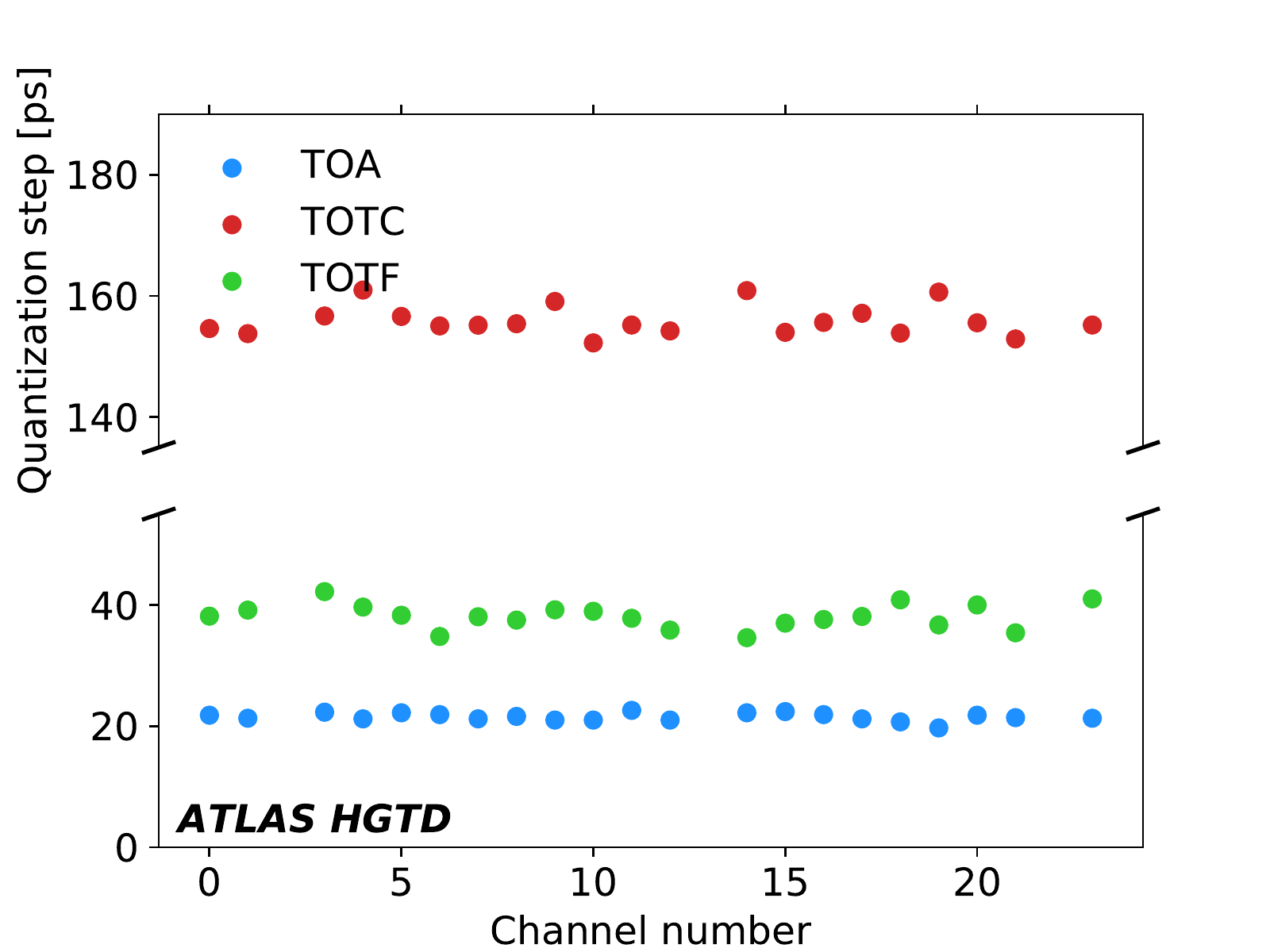}\label{fig:lsb}}
\caption{(a) TOA as a function of the delay together with a linear fit. (b) Quantization step as a function of the channel number for TOA, coarse TOT (TOTC) and fine TOT (TOTF). The measurements were performed with the external discriminator for the sensor board.}
\end{figure}

The first step towards the evaluation of the full single-channel readout is the calibration of the TDC since the TDC quantization steps are needed to obtain the real values of the TOA and TOT.
This was achieved by sending a square external trigger pulse directly to the TDC inputs, the delay of which was adjustable in 10~ps steps. This bypassed the preamplifier and discriminator, allowing direct measurements of the TOA as a function of the delay, as shown in Figure ~\ref{fig:lsbtoa}.
The measured TOA TDC quantization step was found to be around \SI{21.4}{\pico\second}, slightly above the nominal value of \SI{20}{\pico\second}.
As a consequence, the maximum TOA that could be converted was slightly larger than the nominal window of \SI{2.5}{\nano\second}. 
The uniformity of the quantization step for the TOA is shown on Figure ~\ref{fig:lsb} and is better than 3\%.
Measurements are missing for a few channels due to an issue with the SRAM which will be solved for the next ASIC version.
The external trigger had a variable width, adjustable in 10 ps steps which was used to measure the quantization steps for the TOT.
The averaged measured quantization step is \SI{155.2}{\pico\second} (\SI{38.1}{\pico\second}) for the coarse (fine) TOT, close to the nominal value of \SI{160}{\pico\second} (\SI{40}{\pico\second}).
The measured dispersions are better than 2\% (5\%), as can be seen in Figure~\ref{fig:lsb}.
All these results are already close
to the nominal values and they can be further improved using internal TDC slow control parameters to adjust the quantization steps for each channel individually. These further studies are outside the scope of this paper.

\begin{figure}[htbp]
\centering
\subfloat[]{\includegraphics[width=0.5\textwidth]{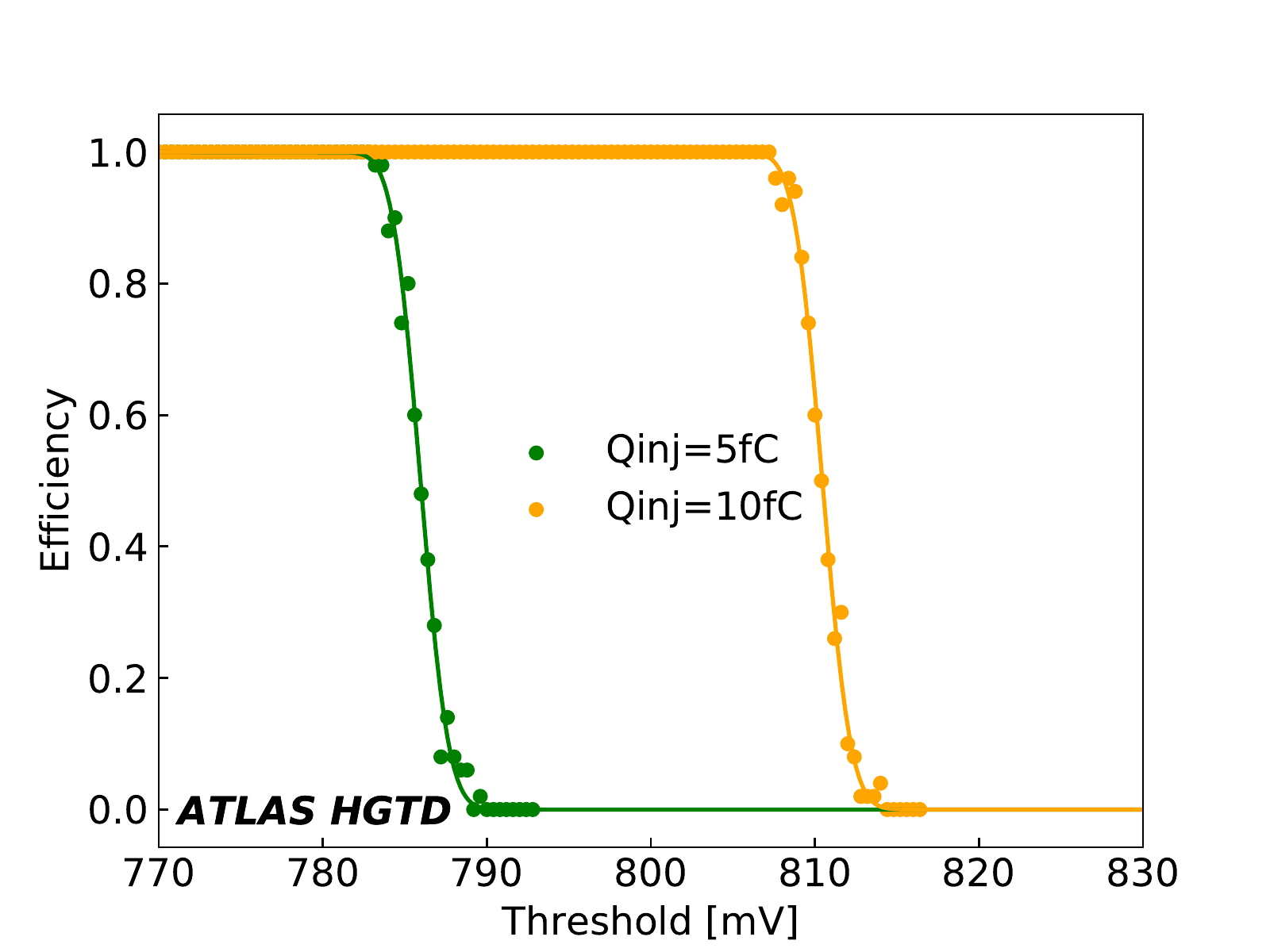}\label{fig:scurve}}
\hspace*{0.2cm}
\subfloat[]{\includegraphics[width=0.5\textwidth]{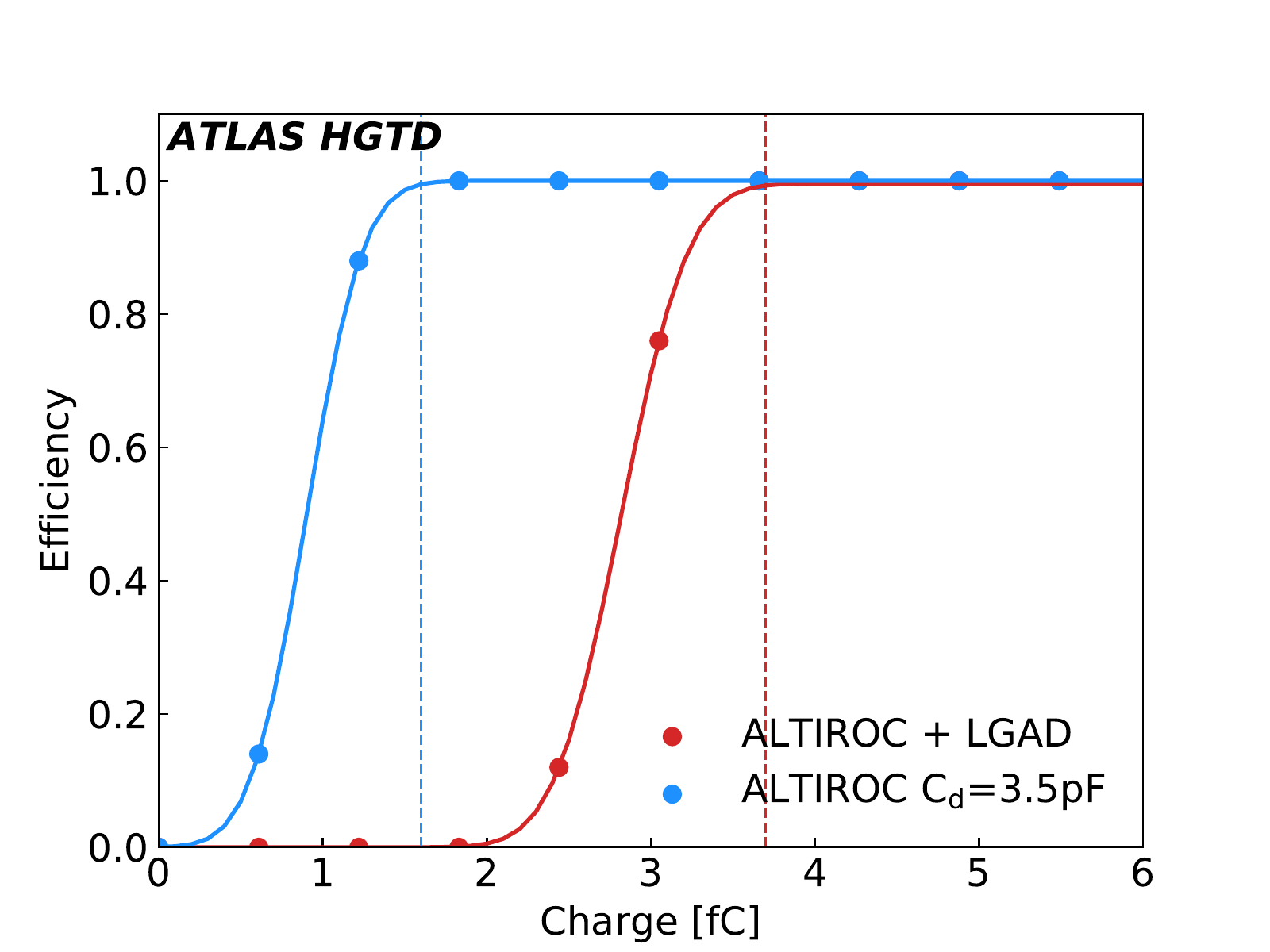}\label{fig:eff}}
\caption{
(a) Efficiency as a function of the discriminator threshold for the ASIC board for a transimpedance preamplifier. The scan automatically stops when the efficiency is 0 for 5 consecutive measurements.
(b) Efficiency as a function of the charge for the lowest possible threshold for transimpedance preamplifiers. For the two figure, the data are fitted with an Gauss error function. The vertical line represents the charge value at which the efficiency is 99\%. 
}
\end{figure}

As described in section~\ref{sec:pulser}, the test bench measurements can also be performed with a $C_{\mathrm{test}}$ capacitor of \SI{200}{\femto\farad} (selectable by slow control) together with a calibration pulser. The pulser generates a Dirac input charge with a relative precision between channels of $\sim 1\%$. The charge is injected at the input of the preamplifier. In this setup, a common 10-bit DAC is used to set the common discriminator threshold for all the channels. The dynamic range goes from 650 mV up to 1 V in steps of 0.4 mV. Using this configuration, the common threshold value was scanned to compute the efficiency, as illustrated in Figure~\ref{fig:scurve} for two different input charges. The efficiency curves were fitted with a Gauss error function:

\begin{equation}
    A\times(1+\erf{ \frac{-(x-\mu)}{\sqrt{2}\sigma}})
\end{equation}

where $A$ is a normalisation factor equal to 0.5 when the highest efficiency is 1, $\mu$ and $\sigma$ are the expected value and the square root of the variance of the related Gaussian function, respectively. The parameter $\sigma$ is an estimator of the voltage noise (N) of the preamplifier output. Since the signal generated by the internal pulser is synchronous with the 40~MHz clock, this noise measurement is not sensitive to any digital coupling synchronized with the clock. 
For the ASIC board, the noise was measured to be 1.0$\pm$0.1~mV for transimpedance preamplifiers and 0.7$\pm$0.1~mV for voltage preamplifiers. The noise for the sensor board is larger than that of the ASIC board by around 20\%. The preamplifier's gain was computed as the difference of the Gauss error function expected values for two input charges divided by the difference of the injected charges. For the transimpedance preamplifiers the gain was measured to be 4.9$\pm$0.2~mV/fC irrespective of the presence of an LGAD sensor array.
For the voltage preamplifiers, the gain is similar to the transimpedance preamplifiers when there is no sensor array (4.7$\pm$0.2~mV/fC). However, the gain is $\sim$20\% lower in the presence of an sensor array (3.7$\pm$0.2~mV/fC). This result is not understood. All these numbers are summarized in table~\ref{tab:properties} together with other quantities characterising the ASIC performance described below.

\begin{table}
\begin{center}
\begin{tabular}{l|cc|cc|}

        \cline{2-5}
        & \multicolumn{2}{c|}{Transimpedance preamp.} & \multicolumn{2}{c|}{Voltage preamp.}\\
        \cline{2-5}
        & ALTIROC & ALTIROC & ALTIROC & ALTIROC\\
        &         &  +LGAD  &         & +LGAD  \\
        \hline
                \multicolumn{1}{|c|}{N [mV]} & \multicolumn{1}{c}{1.0$\pm$0.1} & \multicolumn{1}{c|}{1.2$\pm$0.1}& \multicolumn{1}{c}{0.7$\pm$0.1} & \multicolumn{1}{c|}{0.8$\pm$0.1}\\
     
                     \multicolumn{1}{|c|}{Gain [mV/fC]} & \multicolumn{1}{c}{4.9$\pm$0.2} & \multicolumn{1}{c|}{4.9$\pm$0.2}& \multicolumn{1}{c}{4.7$\pm$0.2} & \multicolumn{1}{c|}{3.7$\pm$0.2}\\
                     
                \multicolumn{1}{|c|}{dV/dt (Q$_{inj}~$=10fC) [mV/ps]} & \multicolumn{1}{c}{0.18$\pm$0.01} & \multicolumn{1}{c|}{0.16$\pm$0.01}& \multicolumn{1}{c}{0.14$\pm$0.01} & \multicolumn{1}{c|}{0.11$\pm$0.01}\\   
                
                \multicolumn{1}{|c|}{N/(dV/dt) (Q$_{inj}~$=10fC) [ps]} & 
                \multicolumn{1}{c}{13$\pm$1} & \multicolumn{1}{c|}{18$\pm$1}& \multicolumn{1}{c}{13$\pm$1} & \multicolumn{1}{c|}{18$\pm$1}\\   
                
                \multicolumn{1}{|c|}{Q$_{min}$ [fC]} & \multicolumn{1}{c}{1.5$\pm$0.1} & \multicolumn{1}{c|}{3.4$\pm$0.1}& \multicolumn{1}{c}{1.5$\pm$0.1} & \multicolumn{1}{c|}{3.4$\pm$0.1}\\  
                
                 \multicolumn{1}{|c|}{jitter  (Q$_{inj}~$=10fC) [ps]}&
                \multicolumn{1}{c}{15$\pm$1} & \multicolumn{1}{c|}{19$\pm$1}& \multicolumn{1}{c}{15$\pm$1} & \multicolumn{1}{c|}{17$\pm$1}\\ 
               
        \hline
      
  \end{tabular}
   
   \caption{\label{tab:properties} Summary of the main averaged ALTIROC signal and noise parameters for transimpedance and voltage preamplifiers and for the ASIC and sensor boards.}
\end{center}
\end{table}

The discriminator threshold was also scanned to find the lowest detectable charge (Q$_{min}$) which is obtained for the lowest discriminator threshold with no data and when no charge is injected.
Figure~\ref{fig:eff} shows the efficiency as a function of the charge for the lowest discriminator threshold. The data were fitted with a Gauss error function, and this fit was used to extract the lowest detectable charge (defined as the charge value corresponding to an efficiency of 99\% and represented as a vertical dashed line on the figure).
For the ASIC board, the lowest detectable charge was found to be on average 1.5$\pm$0.1~fC regardless of the preamplifier type, with a dispersion of 0.2~fC. These results match the ASIC requirement. However, when the ASIC was connected to an LGAD sensor array, the lowest detectable charge increased to 3.4$\pm$0.1~fC, regardless of the preamplifier type, with a dispersion of 0.3~fC. The degraded performance is attributed to a digital coupling synchronized with the 40 MHz clock. The digital activity at the channel level generates a 40 MHz digital noise, not included in the noise measurements in table~\ref{tab:properties}, that is partially injected through the ground of the preamplifier and so amplified by it. The floorplan has been improved for the next version of the ASIC to decrease this effect.

\begin{figure}[htbp]
\centering
\subfloat[]{\includegraphics[width=0.5\textwidth]{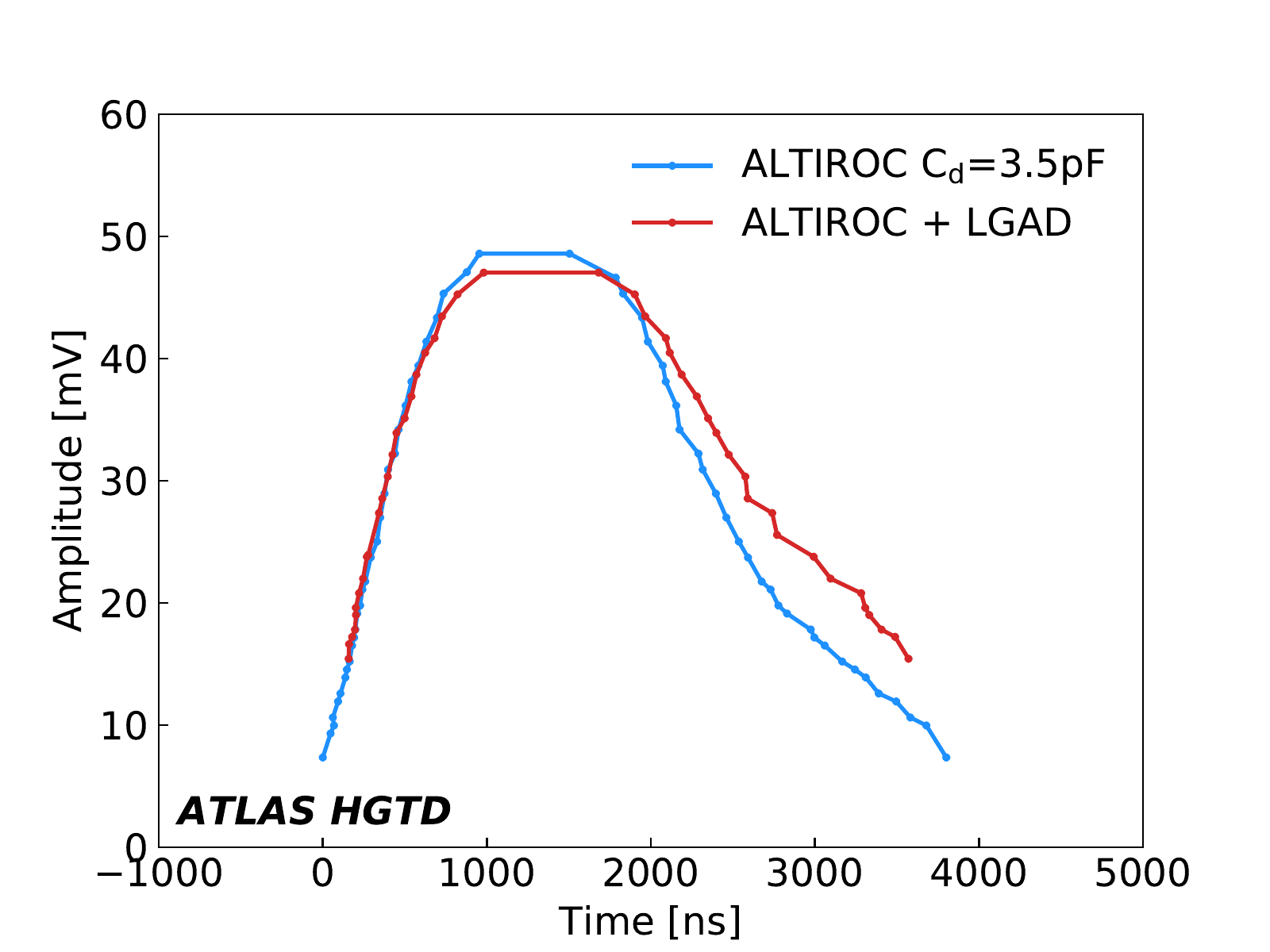}\label{fig:PS_TZ}}
\hspace*{0.2cm}
\subfloat[]{\includegraphics[width=0.5\textwidth]{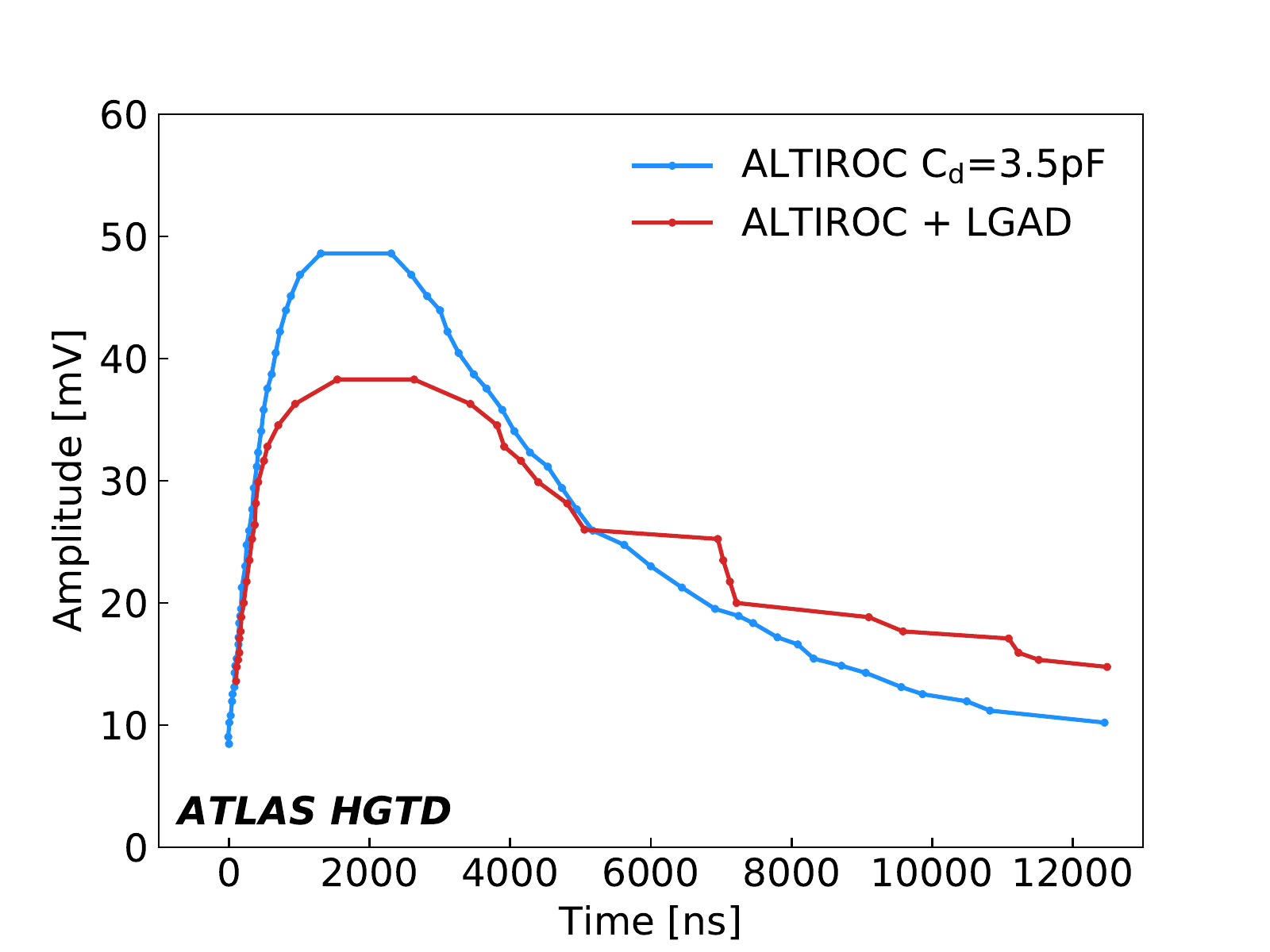}\label{fig:PS_VPA}}
\caption{
Reconstructed analog pulse using timing information from threshold scans for transimpedance (a) and voltage (b) preamplifiers.
\label{fig:TestBenchPS}
}
\end{figure}

The threshold scans were used to reconstruct the analog pulse shape using the information of the TDC by computing the average TOA and TOT for each value. Figures \ref{fig:PS_TZ} and \ref{fig:PS_VPA} show the reconstructed pulse shape for transimpedance and voltage preamplifiers respectively, for an injected charge of 10~fC. In each case, the results with and without an LGAD sensor array are compared. First of all, the pulse duration is significantly lower for the transimpedance preamplifier compared to the voltage preamplifier, as expected (see figure~\ref{fig:simu1}).
The presence of the LGAD sensor array alters the pulse shape, particularly the falling edge, due to couplings at the entrance of the preamplifier. This effect is more prominent for the voltage preamplifiers and has significant implications for the TOT measurements, as described below. The rising edge is not affected by this issue and the pulse derivative ($dV/dt$) is extracted using a linear fit (see table~\ref{tab:properties}).
The pulse derivative is larger for the transimpedance preamplifiers compared to the voltage preamplifiers, compensating for a larger noise. The expected jitter computed as $N/(dV/dt)$ is identical for the two types of preamplifiers for the ASIC board and is equal to 13$\pm$1~ps. The expected jitter is slightly larger in the presence of an LGAD sensor array and is equal to 18$\pm$1~ps.

\begin{figure}[htbp]
\centering
\subfloat[]{\includegraphics[width=0.5\textwidth]{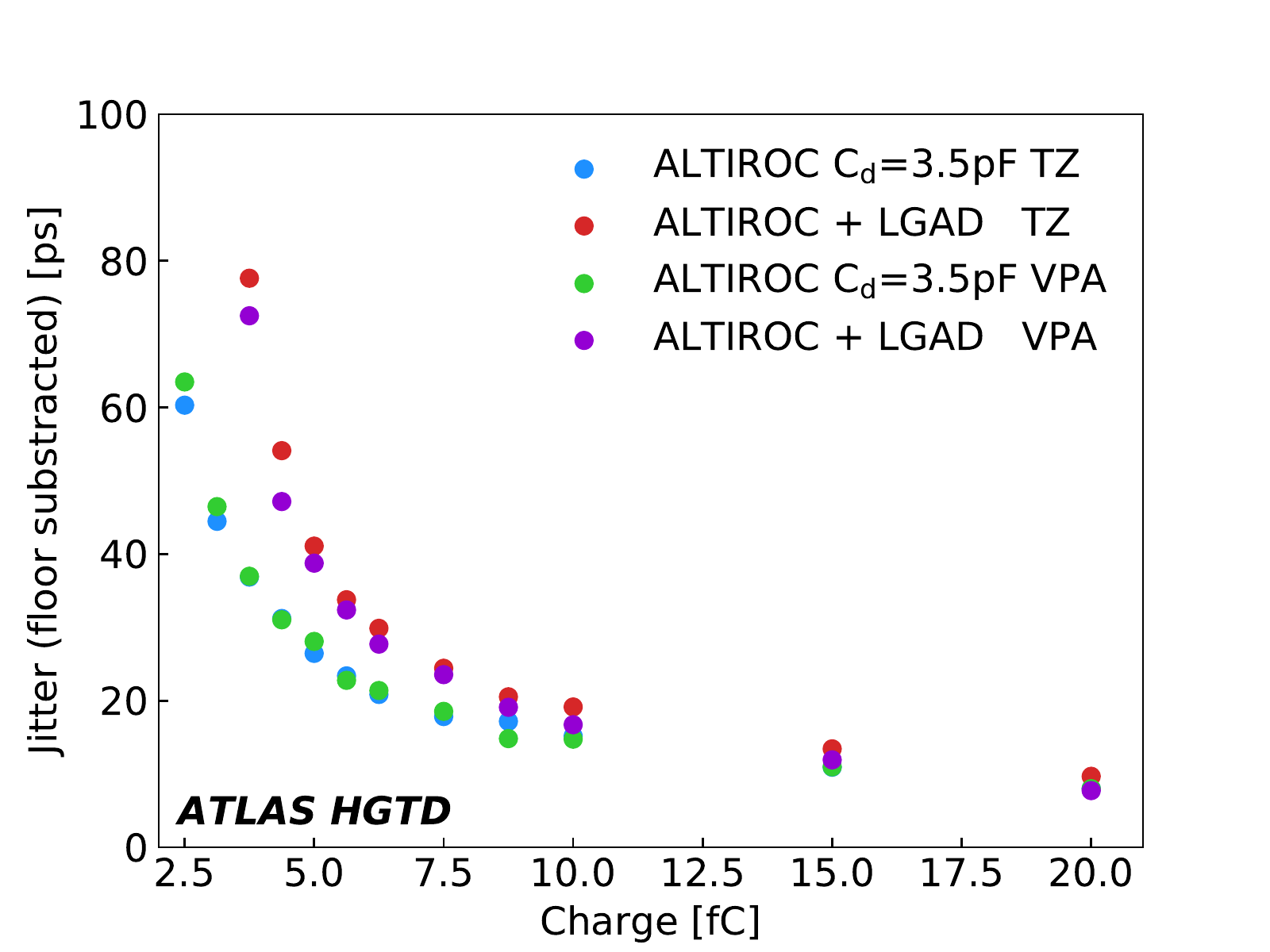}\label{fig:jitter}}
\hspace*{0.2cm}
\subfloat[]{\includegraphics[width=0.5\textwidth]{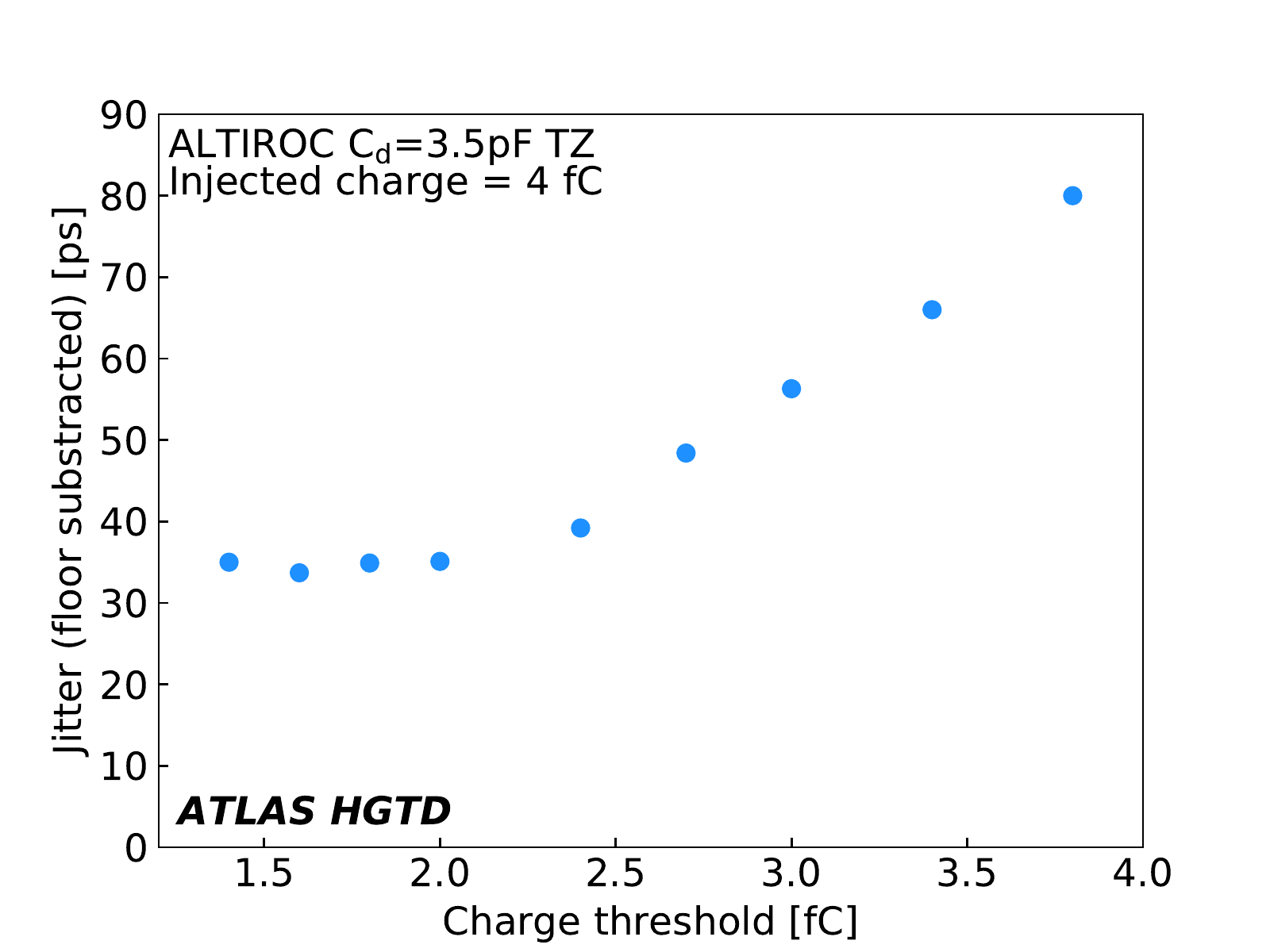}\label{fig:jitter_vs_thres}}
\caption{(a) Jitter averaged over all channels as a function of the injected charge after floor substraction. (b) Jitter averaged over all channels as a function of the charge threshold for the ASIC board using a transimpedance preamplifier.}
\end{figure}

The jitter can be calculated from the RMS of the TOA distribution.
Figure \ref{fig:jitter} shows the jitter variation as a function of the input charge for
the ASIC and sensor boards.
In the case of the ASIC board, the discriminator threshold was set to 2~fC while for the sensor board, the lowest possible threshold was used (Q$_{min}$=3.4~fC). 
The jitter is averaged over all channels separately for the two preamplifier types. The contribution to the jitter, which is independent of the charge due to the clock distribution and the TDC, was subtracted from the measurements. This contribution was obtained from the width of the TOA distribution for the highest injected charge ($\sim$40~fC) and was found to be about 15~ps.
The jitter is very similar for transimpedance and voltage preamplifiers but differences were observed with and without the sensor array, in particular at low charge.
For an injected charge of 10~fC, the jitter is 15$\pm$1~ps for the two preamplifier types for the ASIC board. For the sensor board, the jitter is 19$\pm$1~ps (17$\pm$1~ps) for transimpedance (voltage) preamplifiers. The values agree with the prediction using the measured noise and the derivative of the pulse, as shown in table~\ref{tab:properties}. The results satisfy the ASIC specification. The reasonable agreement with and without the sensor array confirms that the 3.5~pF capacitor connected to the preamplifier correctly mimic the presence of an LGAD array.


For an injected charge of 4~fC, the jitter was measured to be 35$\pm$1~ps for the ASIC board regardless of the preamplifier type. The value scaled reasonably well with the expected charge dependence of the jitter. However, when the sensor array was attached, the jitter significantly degraded. 
This effect was traced back to a non-optimal value of the discriminator threshold utilized in the measurements, due to the aforementioned digital coupling.
Figure~\ref{fig:jitter_vs_thres} shows that the jitter for an injected charge of 4~fC increases with the charge threshold for the ASIC board using a transimpedance preamplifier. Since the jitter is proportional to the pulse derivative and that the pulse derivative reaches a maximum at half of the signal amplitude, the jitter is expected to be minimal for a threshold of $\sim$2fC. The jitter is also expected to increase at lower thresholds, but measurements with a threshold below 1.5~fC are not possible due to the noise.
The jitter of the ASIC board was measured using the same thresholds as those used for the sensor board, and similar jitter levels were observed. These results confirm that the difference in jitter can be attributed to the difference in the discriminator thresholds between the two boards.
 Even though the jitter is degraded with the presence of the LGAD sensor array, the values are still in agreement with the specification (below 65~ps).

\begin{figure}[htbp]
\centering
\subfloat[]{\includegraphics[width=0.5\textwidth]{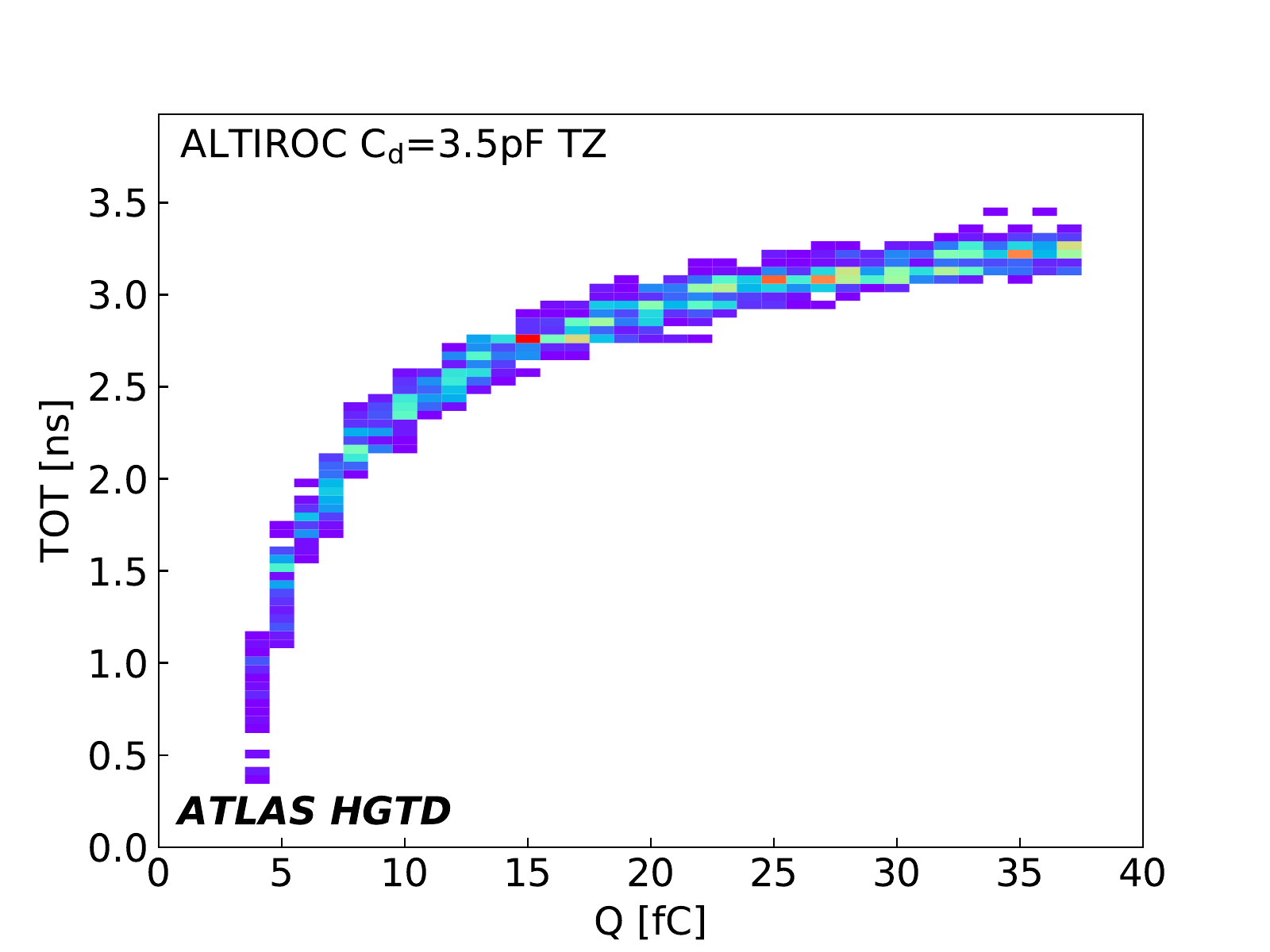}\label{fig:TOT_ALTIROC_TZ}}
\hspace*{0.2cm}
\subfloat[]{\includegraphics[width=0.5\textwidth]{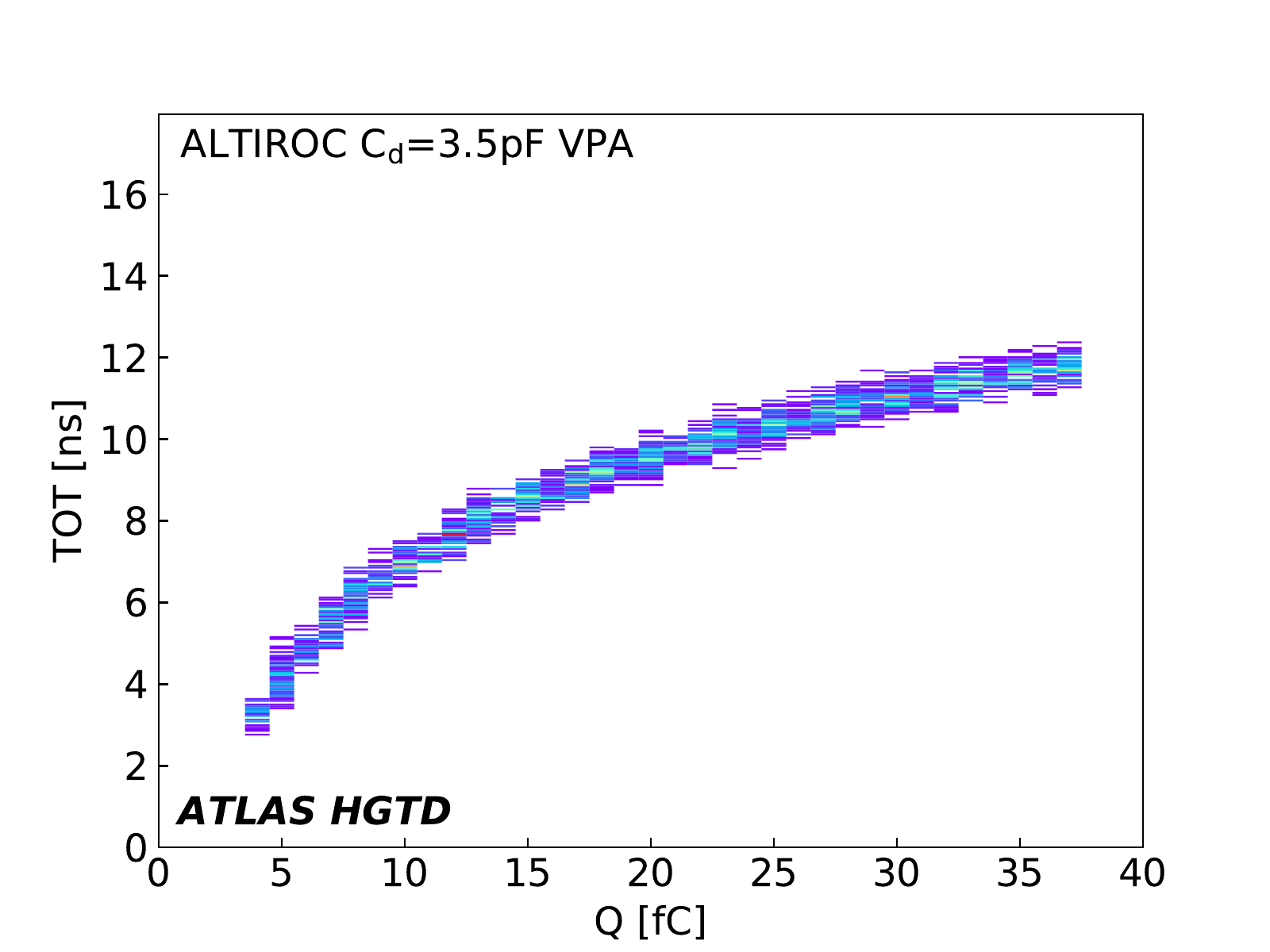}\label{fig:TOT_ALTIROC_VPA}}\\
\subfloat[]{\includegraphics[width=0.5\textwidth]{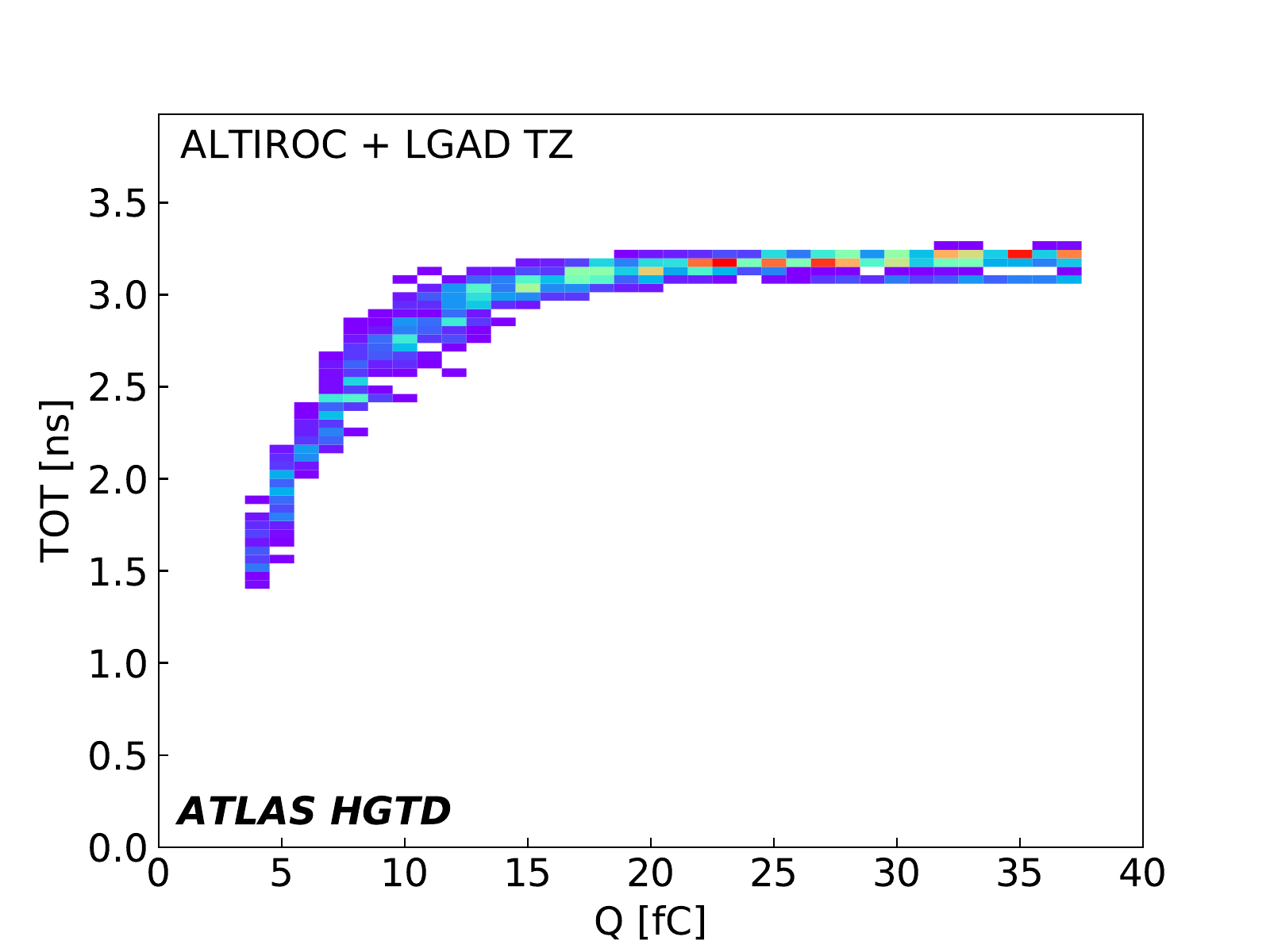}\label{fig:TOT_ALTIROCLGAD_TZ}}
\hspace*{0.2cm}
\subfloat[]{\includegraphics[width=0.5\textwidth]{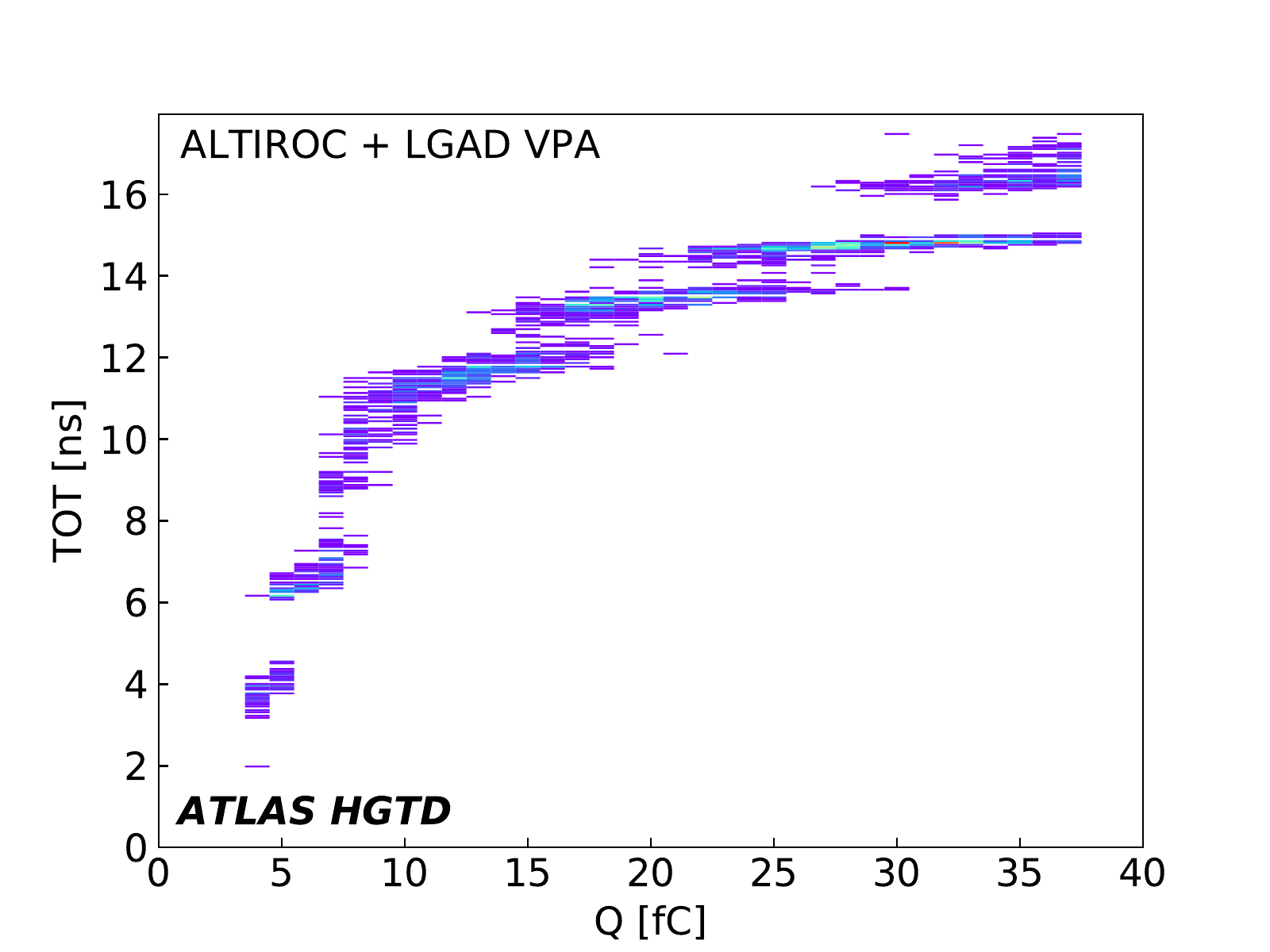}\label{fig:TOT_ALTIROCLGAD_VPA}}
\caption{Distribution of the TOT as a function of the injected charge for the ASIC board (top) and the sensor board (bottom) for a transimpedance (left) and voltage (right) preamplifier. \label{TOT}}
\end{figure}

To correct for time-walk effect, the TOT is used to estimate the signal charge.
Figure~\ref{TOT} shows the TOT (computed with a threshold equal to $\sim$3.4~fC) as a function of the injected charge.
For the ASIC board, the variation of the TOT with the charge is smooth and, as expected, the TOT range is larger for the voltage preamplifier compared to the transimpedance preamplifier. 
When the ASIC is connected to an LGAD sensor array, the TOT measurements are perturbed by the couplings at the input of the preamplifiers, as shown in figure~\ref{fig:simu1}.
For the voltage preamplifier, the TOT does not scale proportionally with the injected charge as already observed in reference~\cite{altiroc0}. In this case, the TOT can't be used to correct for time-walk effect.
Given its shorter pulse, the transimpedance preamplifier is less sensitive to this issue. However, for a charge larger than 15~fC, the TOT has a weak dependence on the input charge, making the time-walk correction less efficient. The next section compares different methods to perform the time-walk correction using beam test data.

\section{Beam test results}
\label{sec:testbeam}
\subsection{ Data analysis}
The ALTIROC1 sensor board was exposed to 120 GeV pions 
at the H6B beamline at the CERN-SPS North Area during two weeks in Autumn 2021.
The setup was similar to the one used in previous HGTD beam tests~\cite{altiroc0,hgtd_tb_paper,hgtd_tb_paper_2018,hgtd_tb_paper_2022} and the differences are mentioned below.
The setup included a EUDET telescope~\cite{telescope} based on six MIMOSA pixel planes, combined with an FE-I4~\cite{fei4} readout chip-based module. The trigger logic was handled by a programmable Trigger Logic Unit (TLU)~\cite{eudettlu}. The telescope was only used for the alignment.

The timing reference was provided by two Cherenkov counters consisting
of a 6$\times$6 mm$^2$ quartz bar coupled to a 5$\times$5 mm$^2$ silicon photomultiplier (SiPM).
analog ASIC probe were sampled by a Lecroy oscilloscope with  sampling rate of 20 GSamples/s and a bandwidth of 2.5 GHz. The 40 MHz clock was also sampled to be able to associate the digital information to the oscilloscope pulses.

A trigger was issued by ALTIROC when a hit was recorded, initiating the data acquisition of the oscilloscope.
While the digital data
acquisition was done continuously, the oscilloscope buffered the data until its memory was full.
At this point it was necessary to pause the data acquisition and transmit the oscilloscope data. Therefore, a ``busy'' logic was implemented in
the FPGA to stop the ASIC acquisition while the oscilloscope was read out.
A common event number provided a unique mapping between the two data streams.


A complete description of the oscilloscope waveform processing can be found in Ref.~\cite{hgtd_tb_paper_2018}. The maximum of the pulse amplitude was estimated as the sample with the maximum amplitude. The TOA for each quartz+SiPM system was calculated using the constant fraction discriminator (CFD) method: it is defined as the point at which the signal crosses 20~\% of its maximal amplitude. The value was chosen to minimize the time resolution.
Three independent time measurements were available, allowing the time resolution for each device to be extracted. The first measurement is the one provided by the sensor board and the two others are the ones from the quartz+SiPM systems. The times are measured with respect to the 40~MHz clock. Three time differences can be computed and fitted individually with a Gaussian function to extract the root mean square width.
For each pair, this Gaussian width is equal to the quadratic sum of the time resolution of each device. This gives a linear system of three equations and three unknowns which can be solved analytically.
The time resolution of the sensor board extracted with this method is the quadratic sum of the Landau fluctuations on the signal shape, the ALTIROC jitter, the clock jitter and the time-walk residual effect.

All the measurements were performed at room temperature.
The device was operated close to breakdown (-265 V), resulting in a MIP charge deposit of about 15 fC. 
The ALTIROC discriminator threshold was set to 5~fC.


\subsection{Results}

The time resolutions of the two quartz+SiPM systems were measured to be 43.4$\pm$0.7~ps and 58.6$\pm$0.7~ps. The system with the smallest time resolution is used in the following figures. The time of the quartz+SiPM system is measured with respect to the 40~MHz clock, allowing direct comparisons with the TOA measured by ALTIROC.

As explained in the previous section, the TOT can't be used to correct the time-walk effect for the voltage preamplifiers. Hence, this type of preamplifier is not considered in this section. For transimpedance preamplifiers, various strategies to perform the time-walk are presented and compared.

\subsubsection{TOT measurements}

\begin{figure}[htbp]
\centering
\subfloat[]{\includegraphics[width=0.5\textwidth]{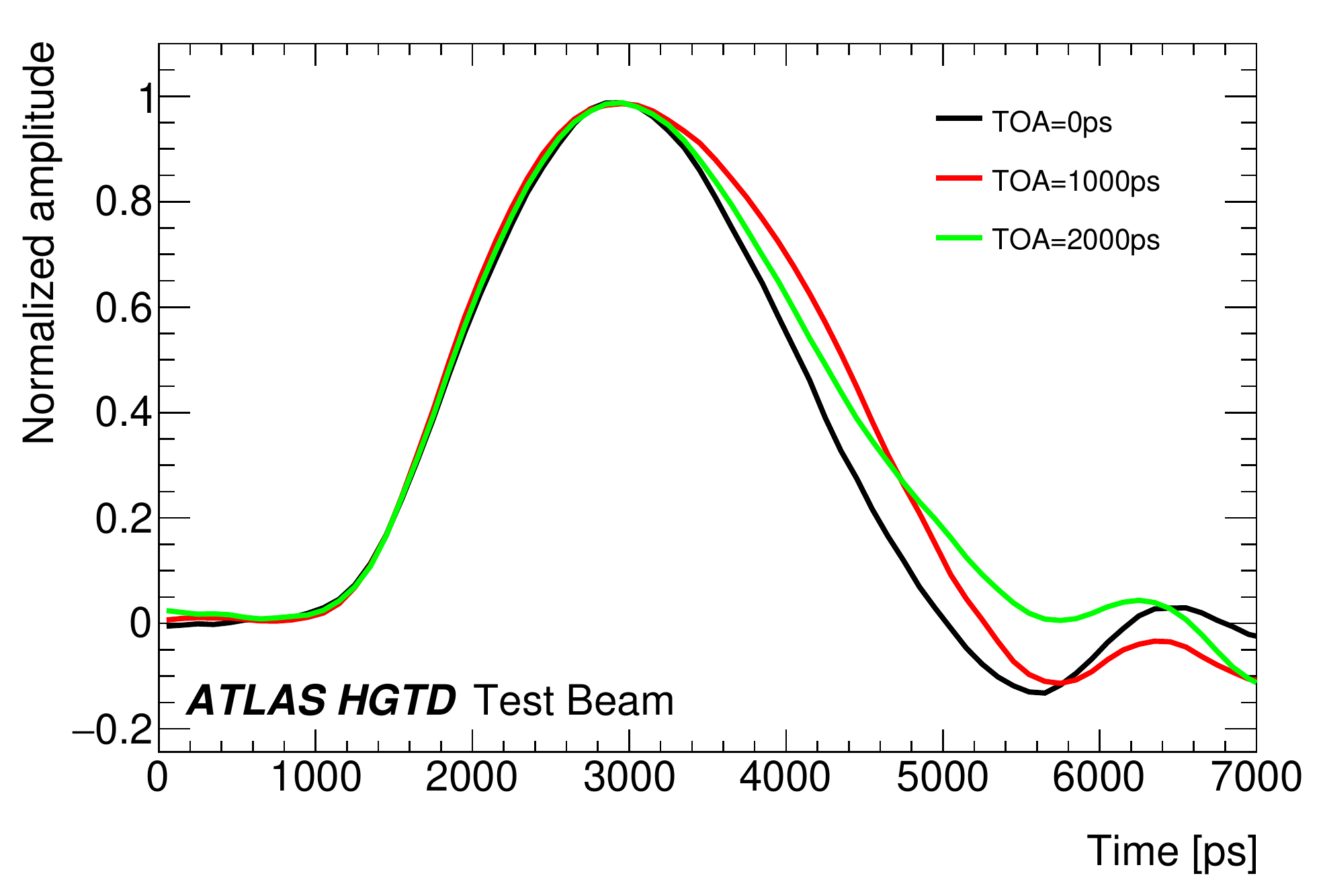}\label{fig:TBPulseShapes}}
\hspace*{0.2cm}
\subfloat[]{\includegraphics[width=0.5\textwidth]{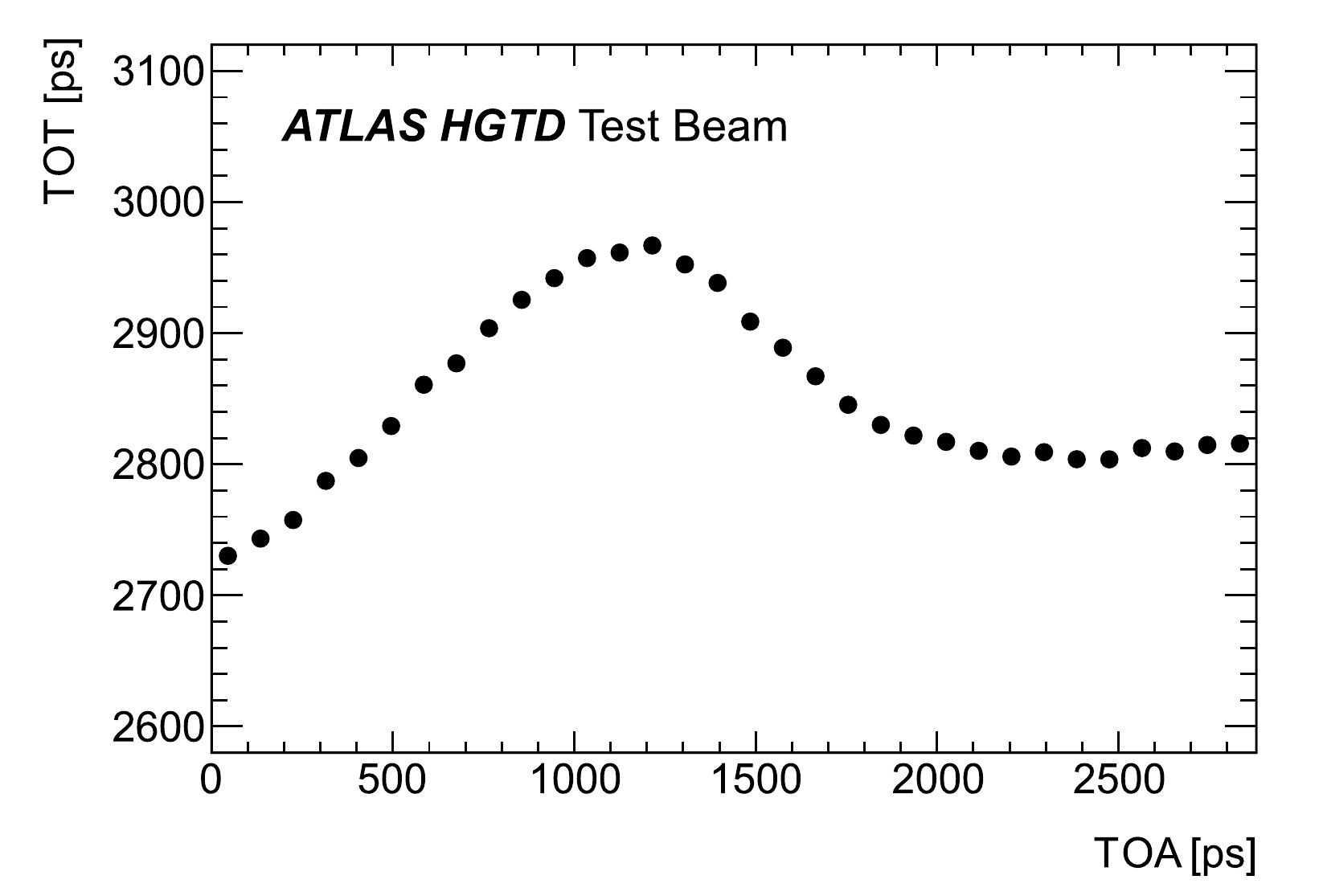}\label{fig:TBTOTTOA}}
\caption{(a)
Preamplifier probe waveform for various TOA values. (b) TOT as a function of the TOA.}
\end{figure}

Figure~\ref{fig:TBPulseShapes} shows the average preamplifier probe signal shape, normalized to unity, for various values of the TOA. The pulse shape depends on the TOA of the incoming particles due to the coupling issues mentioned above.
Consequently, the TOT depends on the TOA, as shown in figure~\ref{fig:TBTOTTOA}. 
The smaller the TOA, the closer it is to the 40~MHz clock edge. For TOA values greater than 1600~ps, the TOT becomes almost independent of the TOA. 

\subsubsection{Time-walk correction using preamplifier probe information}

\begin{figure}[htbp]
\centering
\subfloat[]{\includegraphics[width=0.5\textwidth]{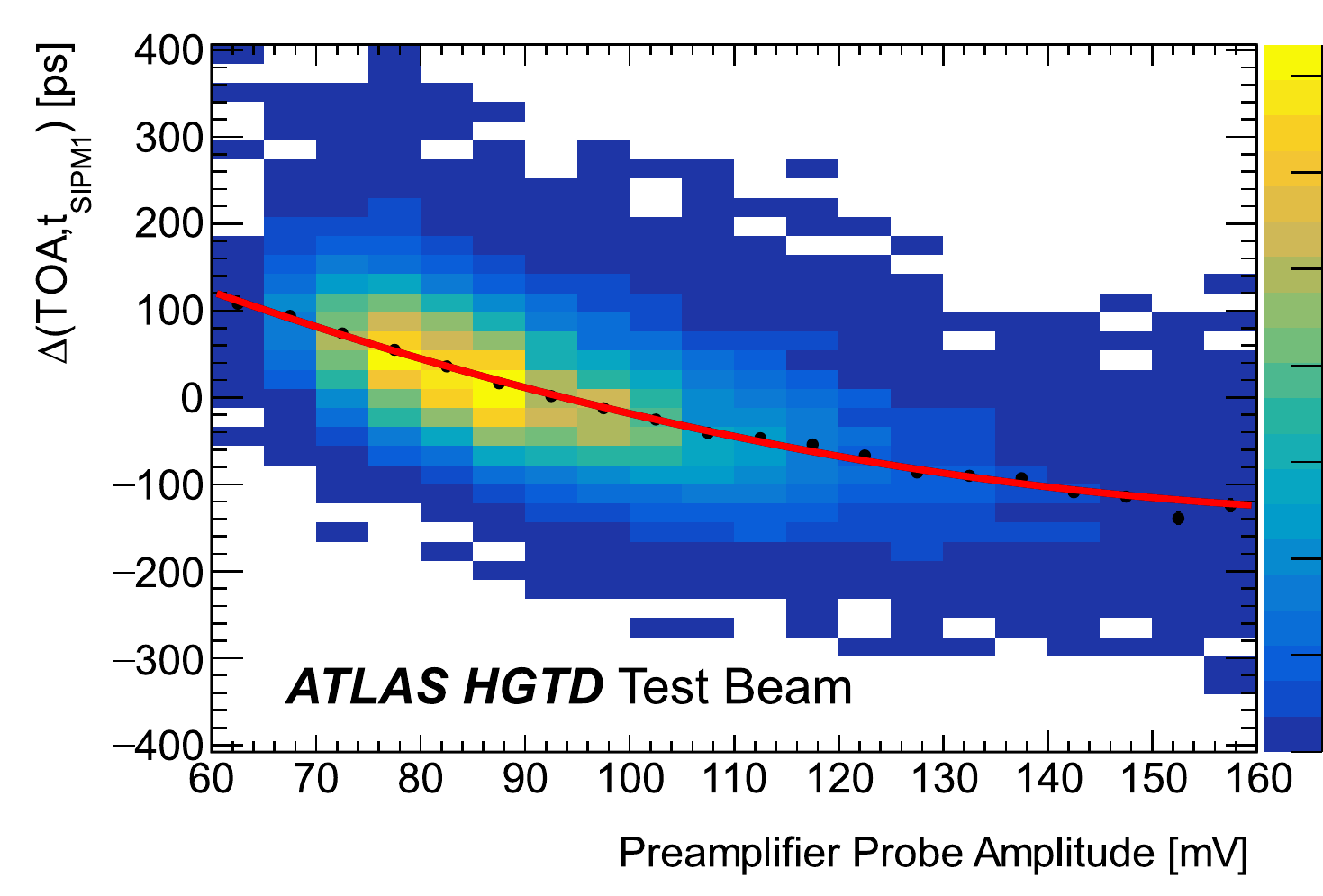}\label{fig:TWprobe}}
\hspace*{0.2cm}
\subfloat[]{\includegraphics[width=0.5\textwidth]{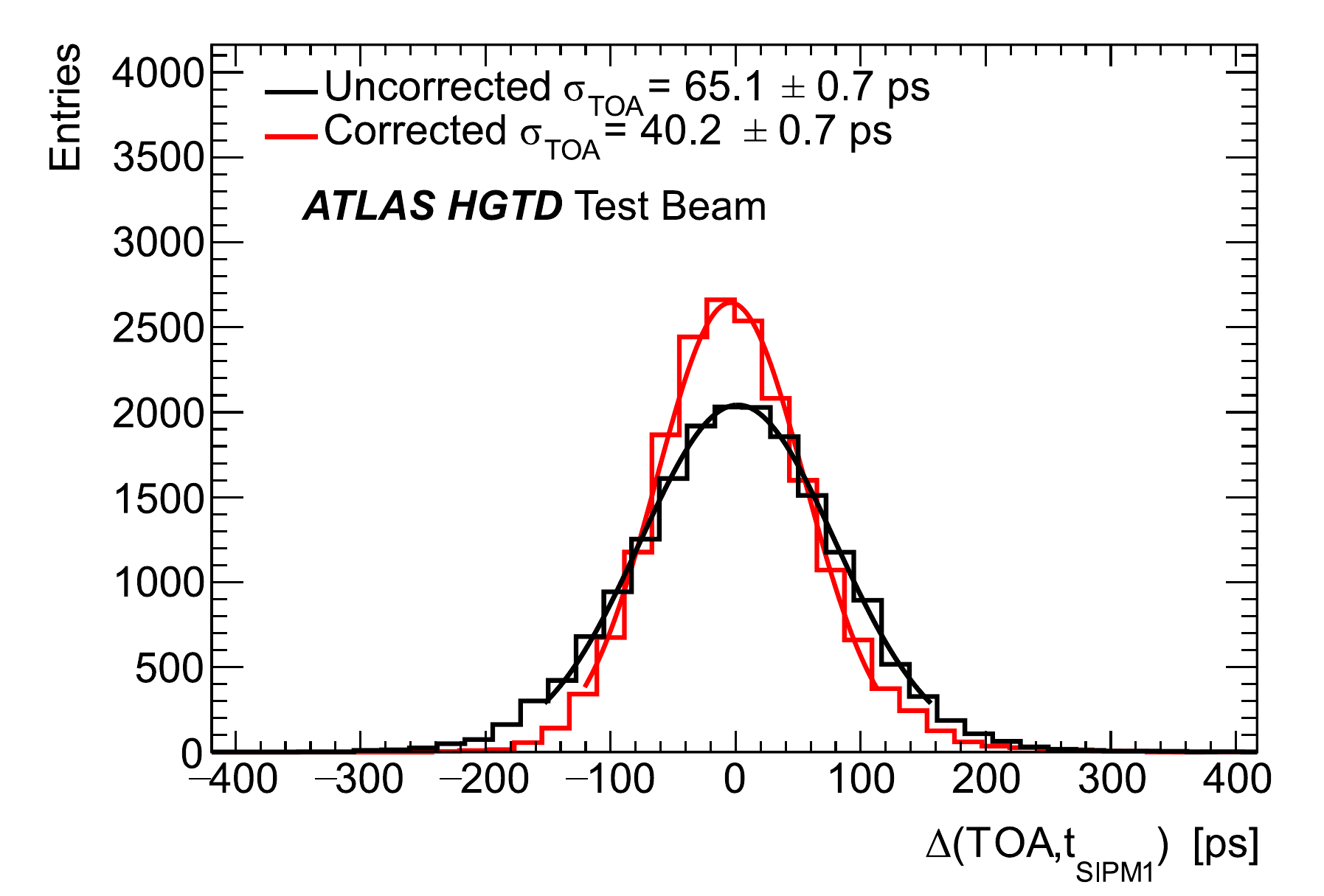}\label{fig:DeltaTprobe}}
\caption{(a)
Distribution of the time difference between the sensor board and quartz+SiPM system as a function of the preamplifier probe amplitude. The dots correspond to the mean value of the TOA distribution and the red line is a
fit used to perform the time-walk correction. (b) Distributions of the time difference between
the sensor board and quartz+SiPM systems before (red) and after (black) the time-walk correction
together with Gaussian fits. The numbers are the fitted Gaussian widths where the time resolution of
the quartz+SiPM system has been subtracted quadratically.
\label{fig:probe}
}
\end{figure}

Figure~\ref{fig:TWprobe} shows time difference between the TOA measured by the sensor board and that from the quartz+SiPM system as a function of preamplifier probe amplitude.  The black dots display the mean for each preamplifier probe amplitude slice and the red line is a polynomial fit to these points used to perform the time-walk correction. Figure~\ref{fig:DeltaTprobe} shows the distribution of the time difference between the TOA measured by the sensor and that from the quartz+SiPM system before and after time-walk correction. The time resolution of the sensor board is 65.1$\pm$0.7~ps before and 40.2$\pm$0.7~ps after time-walk correction. The time resolution after correction is compatible with the time resolution measured at room temperature for HPK2 sensors with the same doping level with a radioactive $\beta$ source for an input charge of 15~fC, read out with custom electronics boards~\cite{ucsc_readoutboard}. One can conclude that the jitter is the same for the two electronic circuits and that the residual time-walk contribution is negligible.

\begin{figure}[htbp]
\centering
\includegraphics[width=0.5\textwidth]{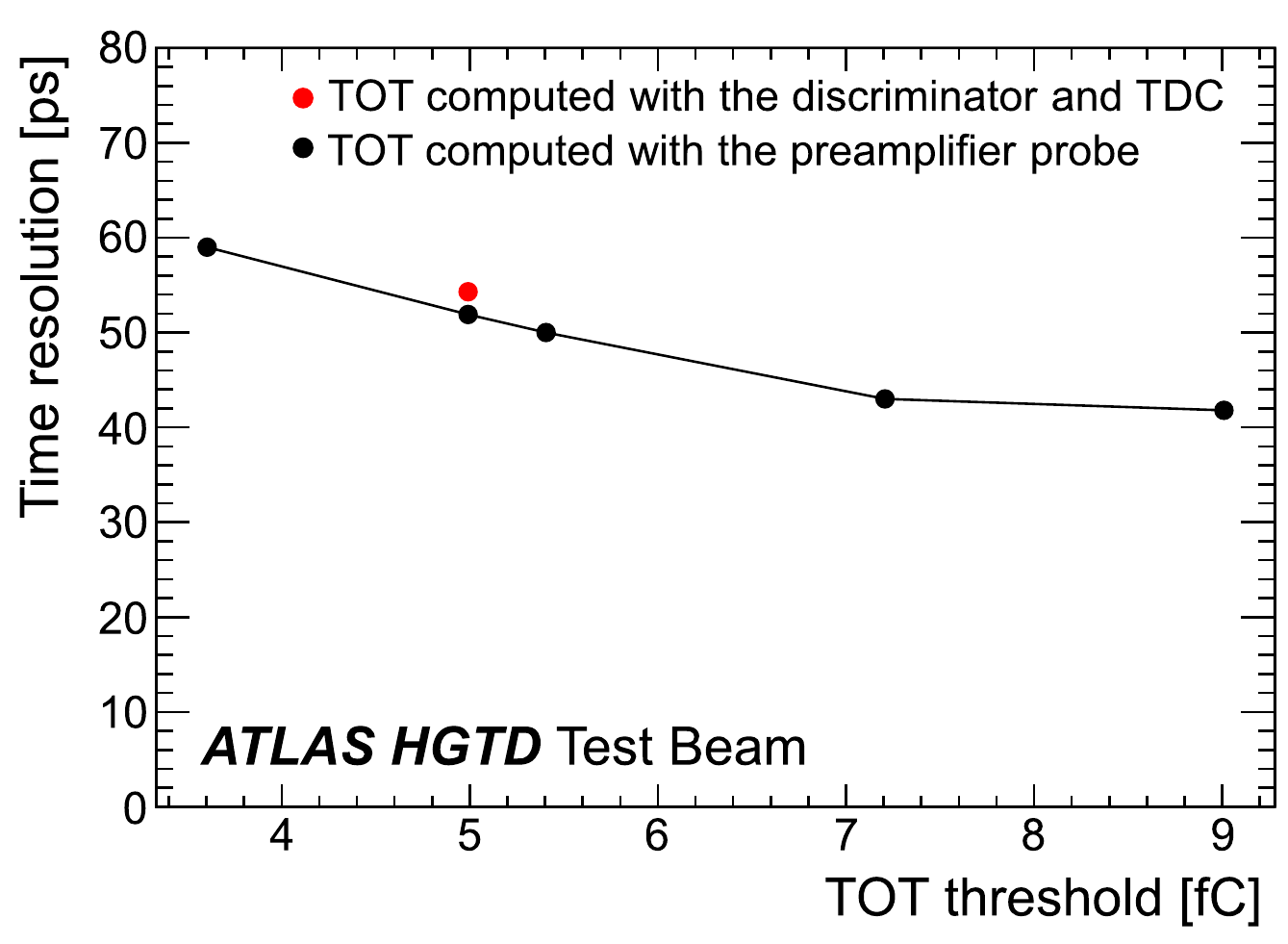}
\caption{
Time resolution after time-walk correction as a function of the probe TOT threshold.\label{fig:probeTOT}}
\end{figure}

The time-walk can also be corrected using the TOT computed from the preamplifier probe, which is a proxy for the signal amplitude. Figure~\ref{fig:probeTOT} shows the variation of the time resolution after time-walk correction as a function of the threshold used to compute the TOT. For the largest threshold, the jitter is similar to the one obtained using the preamplifier probe amplitude for the time-walk correction.
Lowering the thresholds results in an increase in time resolution.
For instance, the time resolution after correction is 55$\pm$0.7~ps for a threshold of 5~fC.
Subtracting the time resolution after correction using the preamplifier probe amplitude gives the residual time-walk contribution which is found to be 37~ps. 
The degradation of the time resolution is attributed to the dependence of the TOT on the TOA, due to the digital coupling being synchronized with the 40 MHz clock.



\subsubsection{Time-walk correction using TDC information}

At the HL-LHC, the HGTD is expected to use the TOT computed from the discriminator and the TDC for the time-walk correction. Using this method for the beam test data, the time resolution after correction was measured to be 54.2$\pm$0.7~ps at a threshold of 5~fC. This result is consistent with the time resolution obtained after time-walk correction using the preamplifier probe TOT.

The time-walk correction was also performed using a modified version of the TOT measured by ALTIROC: for each TOA bin, the average TOT was subtracted from the TOT. This new variable called corrected TOT (corTOT) is by construction independent of the TOA. In this case, the time resolution after correction was found to be 49.6$\pm$0.7~ps. This method helps to reduce the residual time-walk contribution from 36~ps using the default TOT to 29~ps.

\begin{figure}[htbp]
\centering
\subfloat[]{\includegraphics[width=0.5\textwidth]{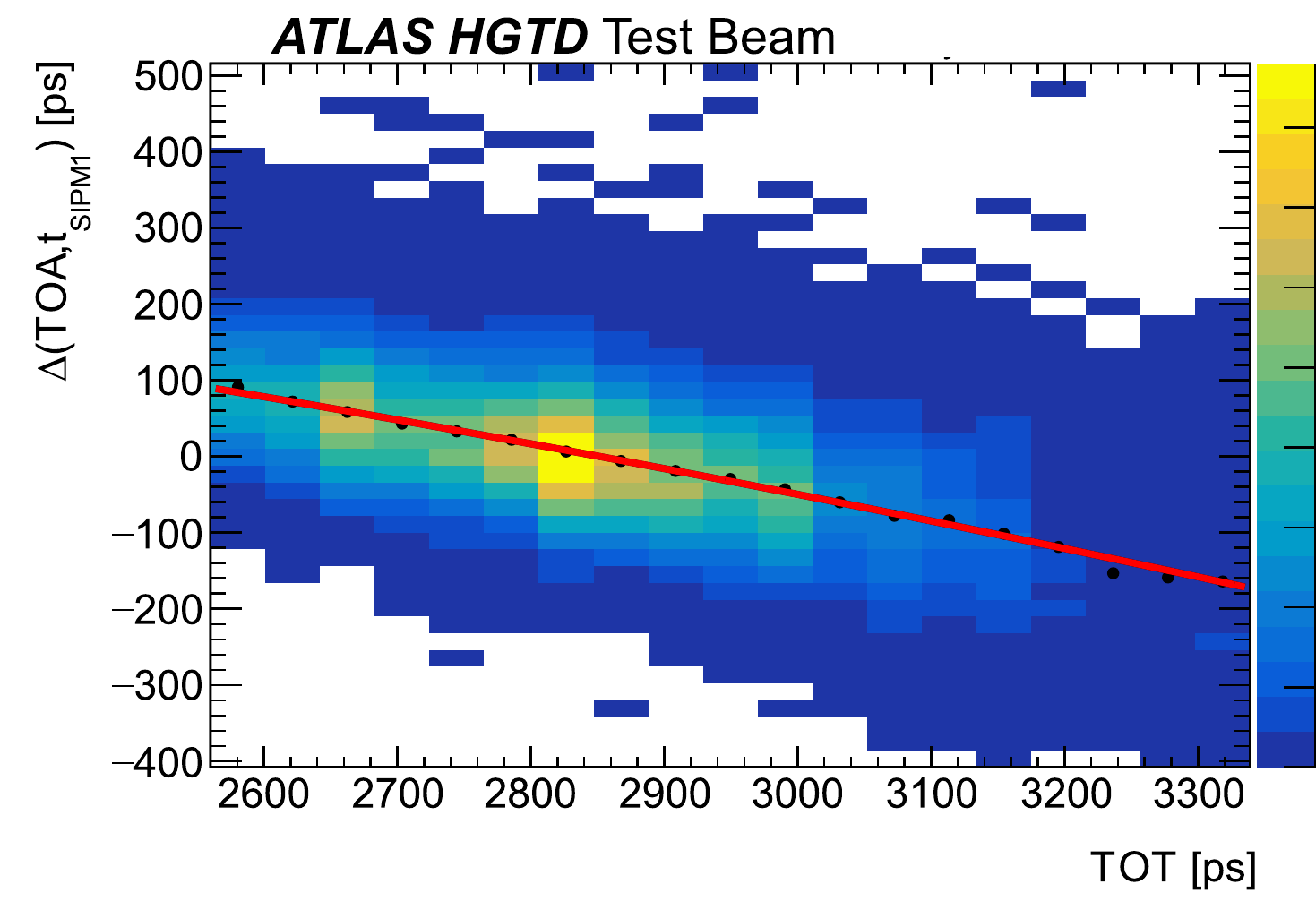}\label{fig:TWTOT}}
\hspace*{0.2cm}
\subfloat[]{\includegraphics[width=0.5\textwidth]{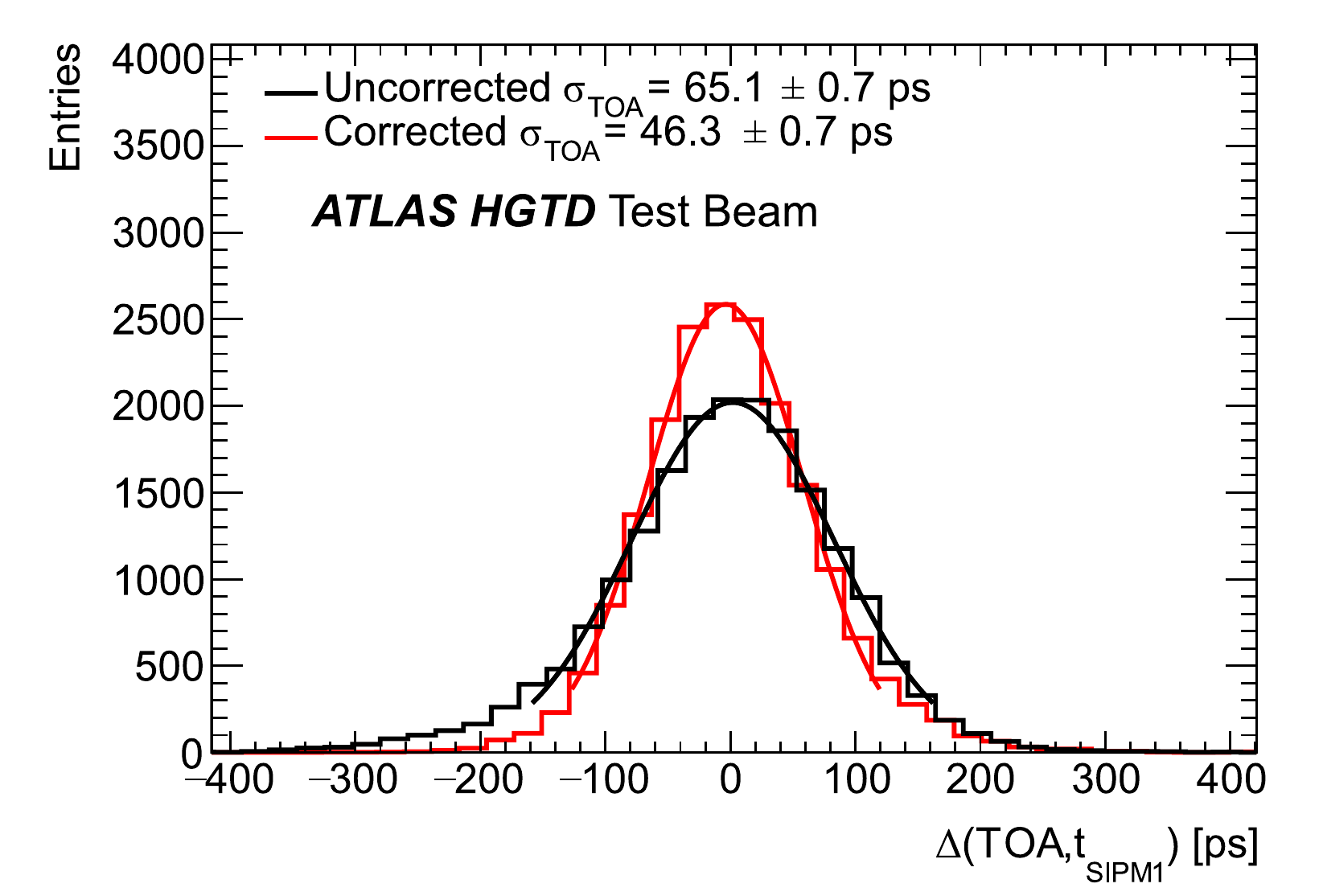}\label{fig:DeltaTTOT}}
\caption{(a)
Distribution of the time difference between
the sensor board and quartz+SiPM system as a function of the TOT, for TOA values greater than 1600~ps. The dots correspond to the mean value of the TOA distribution and the red line is a
fit used to perform the time-walk correction. (b) Distributions of the time difference between
the sensor board and quartz+SiPM systems before (red) and after (black) time-walk correction
together with Gaussian fits for TOA greater than 1600~ps. The numbers are the fitted Gaussian widths where the time resolutions of
the quartz+SiPM system has been subtracted quadratically.
\label{fig:TOT}
}
\end{figure}

Finally, the time resolution was also measured for a restricted range of the TOA values (greater than 1600~ps). In this window of 900~ps, the TOT has a weak dependence on the TOA and in this case the time resolution was measured as 46.3$\pm$0.7~ps (see figure\ref{fig:TOT}) which is the best time resolution obtained using only TDC data from ALTIROC. 
The residual time-walk contribution is equal to 23~ps and remains the dominant electronic contribution to the time resolution, since the jitter contribution is 15~ps at 15~fC as shown in figure~\ref{fig:jitter}. A summary of the time resolution obtained for various time-walk correction methods can be found in table~\ref{tab:testbeam}.

\begin{table}[ht]
    \centering
    \begin{tabular}{|c|c|}
        \hline
        Time-walk correction method & Time resolution  \\
        \hline
        No correction & 65.1$\pm$0.7~ps   \\
        \hline
        Preamplifier probe amplitude &  40.2$\pm$0.7~ps \\
        \hline
        Preamplifier probe TOT &  55.0$\pm$0.7~ps\\
        \hline
        ALTIROC1 TOT &  54.2$\pm$0.7~ps\\
        \hline
        Corrected ALTIROC1 TOT & 49.6$\pm$0.7~ps \\
        \hline
        ALTIROC1 TOT for TOA \textgreater{} 1600~ps & 46.3$\pm$0.7~ps \\
        \hline
          \end{tabular}
    \caption{Time resolution for various time-walk correction methods.}
    \label{tab:testbeam}
\end{table}

\FloatBarrier

\section{Conclusion}
\label{sec:conclusion}
A prototype of the front-end electronics ASIC for the ATLAS High Granularity Timing Detector has been designed and tested with calibration signals and beam test beams. 
This prototype, called ALTIROC1, consists of a 5$\times$5-pad matrix and contains the analog
part of the single-channel readout (preamplifier, discriminator and two TDCs).
Two preamplifier architectures (transimpedance and voltage) have been implemented and tested.

In calibration measurements, the ASIC specification requirements are found to be fully satisfied with ALTIROC1, with similar performances for the two preamplifier options. 
In particular, the jitter is 15$\pm$1~ps (resp. 35$\pm$1~ps) for an injected charge of 10~fC (resp. 4~fC). However, the performance degraded when the ASIC was connected to an LGAD sensor array. This effect is attributed to digital coupling synchronized with the 40 MHz clock.
The lowest detectable charge increased from 1.5~fC to 3.4~fC when the LGAD sensor array was included. While the jitter is almost unchanged at 10~fC, it increases significantly at~4~fC. Nevertheless, the ALTIROC jitter specification (to be below 65~ps) is still met.
The voltage preamplifier is more sensitive to the coupling issue due to its longer pulse duration. The TOT is strongly affected by this effect and can't be used to perform the time-walk correction. The transimpedance preamplifier is less sensitive to this effect thanks to its shorter pulse duration, but the TOT exhibits a dependence with the TOA due to couplings synchronized with the 40~MHz clock.

Beam test measurements with a pion beam at CERN were also undertaken to evaluate the performance of the module using an HPK2 W42 5$\times$5 sensor 
operated at a bias voltage of -265 V. In these conditions, the most probable charge is 15~fC.
The measurements presented in this paper focus only on transimpedance preamplifiers.
A time resolution of 40.2$\pm$0.7~ps was measured when the time-walk correction was performed using the preamplifier probe amplitude. 
The time resolution was worse using the TOT due to the couplings issue. 
The best time resolution obtained using only ALTIROC TDC information was found to be 46.3$\pm$0.7~ps, for a restricted TOA range where the TOT is almost independent of the TOA.
The residual time-walk contribution is equal to 23~ps and is the dominant electronic contribution to the time resolution at 15~fC.

The next iterations of the ASIC will consist of a 15$\times$15-pad matrix, with the introduction of all the digital blocks. In addition to timing measurements, the ASIC will provide the number of hits per bunch crossing for different time windows needed to measure the luminosity.
The design will be optimized to minimize the coupling issue.
Along with the characterisation of the digital blocks of the front-end readout chain, the new ASIC will be evaluated at the HGTD operation temperature and under various irradiation conditions.


\acknowledgments

The authors gratefully acknowledge CERN and the SPS staff for successfully operating the North Experimental Area and for continuous supports to the users.
The USP group acknowledges support from FAPESP grant 2020/04867-2 and CAPES.
This work was partially funded by MINECO, Spanish Government, under grant RTI2018-094906-B-C21. This project has received funding from the European Union's Horizon 2020 research and innovation programme under the Marie Sklodowska-Curie grant agreement No. 754510.


\clearpage


\bibliographystyle{JHEP}
\bibliography{biblio.bib}

\end{document}